%
%
%
%
%
%
%
%
%
%
%
%
\def\firstpage{1}\def\lastpage{1000}
\nopagenumbers

\hoffset=1.66cm\voffset=2.5cm 

\hsize=12.5cm\vsize=19.5cm

\font\caps=cmcsc10                    
\font\Caps=cccsc10 scaled \magstep1   
\font\scaps=cmcsc8

%
\pageno=\firstpage
\def\folio{\rm\number\pageno}\output={\plainoutput}
\def\headfoot{25pt}
\def\makefootline{\baselineskip=\headfoot\line{\the\footline}}
\footline{\hfill\scaps Documenta Mathematica
     $\cdot$ Extra Volume ICM  1998  $\cdot$
    {\ifnum\pageno>\lastpage\else\number\firstpage--\lastpage\fi}\hfill}
\def\DocMath{{\def\th{\thinspace}\scaps Doc.\th Math.\th J.\th DMV}}
\def\makeheadline{
    \vbox to 0pt{\vskip-\headfoot\line{\vbox to8.5pt{}\the\headline}\vss}
    \nointerlineskip}
\headline={\ifnum\pageno=\firstpage{\DocMath\hfill\llap{\folio}}%
           \else{\ifodd\pageno\rightheadline\else\leftheadline\fi}\fi}
\def\rightheadline{\caps \hfill \rightheadtext    \hfill \llap{\folio}}
\def\leftheadline {\caps \rlap{\folio} \hfill \leftheadtext \hfill }
\def\leftheadtext{\ifnum\pageno>\lastpage\else\SAuthor\fi}
\def\rightheadtext{\STitle}
\def\TSkip{\bigskip}
\newbox\TheTitle{\obeylines\gdef\GetTitle #1
\ShortTitle  #2
\SubTitle    #3
\Author      #4
\ShortAuthor #5
\EndTitle
{\setbox\TheTitle=\vbox{\baselineskip=20pt\let\par=\cr\obeylines%
\halign{\centerline{\Caps##}\cr\noalign{\medskip}\cr#1\cr}}%
	\copy\TheTitle\TSkip\TSkip%
\def\next{#2}\ifx\next\empty\gdef\STitle{#1}\else\gdef\STitle{#2}\fi%
\def\next{#3}\ifx\next\empty%
    \else\setbox\TheTitle=\vbox{\baselineskip=20pt\let\par=\cr\obeylines%
    \halign{\centerline{\caps##} #3\cr}}\copy\TheTitle\TSkip\TSkip\fi%
\centerline{\caps #4}\TSkip\TSkip%
\def\next{#5}\ifx\next\empty\gdef\SAuthor{#4}\else\gdef\SAuthor{#5}\fi%
\catcode'015=5}}\def\Title{\obeylines\GetTitle}
\def\Abstract{\begingroup\narrower
    \parskip=\medskipamount\parindent=0pt{\caps Abstract. }}
\def\EndAbstract{\par\endgroup\TSkip}

\long\def\MSC#1\EndMSC{\def\arg{#1}\ifx\arg\empty\relax\else
     {\par\narrower\noindent%
     1991 Mathematics Subject Classification: #1\par}\fi}

\long\def\KEY#1\EndKEY{\def\arg{#1}\ifx\arg\empty\relax\else
	{\par\narrower\noindent Keywords and Phrases: #1\par}\fi\TSkip}

\newbox\TheAdd\def\Addresses{\vfill\copy\TheAdd\vfill
    \ifodd\number\lastpage\vfill\eject\phantom{.}\vfill\eject\fi}
{\obeylines\gdef\GetAddress #1
\Address #2
\Address #3
\Address #4
\EndAddress
{\def\xs{6truecm}
\setbox0=\vtop{{\obeylines\hsize=\xs#1\par}}\def\next{#2}
\ifx\next\empty 
     \setbox\TheAdd=\hbox to\hsize{\hfill\copy0\hfill}
\else\setbox1=\vtop{{\obeylines\hsize=\xs#2\par}}\def\next{#3}
\ifx\next\empty 
     \setbox\TheAdd=\hbox to\hsize{\hfill\copy0\hfill\copy1\hfill}
\else\setbox2=\vtop{{\obeylines\hsize=\xs#3\par}}\def\next{#4}
\ifx\next\empty\ 
     \setbox\TheAdd=\vtop{\hbox to\hsize{\hfill\copy0\hfill\copy1\hfill}
                \vskip20pt\hbox to\hsize{\hfill\copy2\hfill}}
\else\setbox3=\vtop{{\obeylines\hsize=\xs#4\par}}
     \setbox\TheAdd=\vtop{\hbox to\hsize{\hfill\copy0\hfill\copy1\hfill}
	        \vskip20pt\hbox to\hsize{\hfill\copy2\hfill\copy3\hfill}}
\fi\fi\fi\catcode'015=5}}\gdef\Address{\obeylines\GetAddress}

\hfuzz=0.1pt\tolerance=2000\emergencystretch=20pt\overfullrule=5pt
\Title
Chaotic Hypothesis and Universal
Large Deviations Properties
\ShortTitle 
Chaotic Hypothesis \dots
\SubTitle   
\Author 
Giovanni Gallavotti
\ShortAuthor 
\EndTitle
\Abstract 
Chaotic systems arise naturally in Statistical Mechanics and in
Fluid Dynamics. A paradigm for their modelization are smooth
hyperbolic systems. Are there consequences that can be drawn simply by
assuming that a system is hyperbolic? here we present a few model
independent general consequences which may have some relevance for the
Physics of chaotic systems.
\EndAbstract
\MSC 
\EndMSC
\KEY
Chaotic hypothesis, Anosov maps, Reversibility, Large deviations, Chaos
\EndKEY
\Address 
Giovanni Gallavotti
Fisica, Universit\'a La Sapienza
Roma, Italy
giovanni@ipparco.roma1.infn.it
\Address
\Address
\Address
\EndAddress
\newcount\mgnf\newcount\tipi\newcount\tipoformule
\newcount\aux\newcount\piepagina\newcount\xdata
\mgnf=0
\aux=0           
\tipoformule=1   
\piepagina=2     
\xdata=0         
\def\Di{}

\ifnum\mgnf=1 \aux=0 \tipoformule =1 \piepagina=1 \xdata=1\fi
\newcount\bibl
\ifnum\mgnf=0\bibl=0\else\bibl=1\fi
\bibl=0

%
%
%
%
\ifnum\bibl=0
\def\ref#1#2#3{[#1#2]\write8{#1@#2}}
\def\rif#1#2#3#4{\item{[#1#2]} #3}
\fi

\ifnum\bibl=1
\def\ref#1#2#3{[#3]\write8{#1@#2}}
\def\rif#1#2#3#4{}

\fi

\def\9#1{\ifnum\aux=1#1\else\relax\fi}


\let\a=\alpha\let\b=\beta \let\g=\gamma \let\d=\delta
\let\e=\varepsilon \let\z=\zeta \let\h=\eta
\let\th=\vartheta\let\k=\kappa \let\l=\lambda \let\m=\mu 
\let\x=\xi \let\p=\pi \let\r=\rho \let\s=\sigma \let\t=\tau
 \let\f=\varphi  
  \let\D=\Delta 
\let\L=\Lambda   \let\F=\Phi
  
{\count255=\time\divide\count255 by 60 \xdef\oramin{\number\count255}
\multiply\count255 by-60\advance\count255 by\time
\xdef\oramin{\oramin:\ifnum\count255<10 0\fi\the\count255}}
\def\ora{\oramin }

\ifnum\xdata=0
\def\data{\number\day/\ifcase\month\or gennaio \or
febbraio \or marzo \or aprile \or maggio \or giugno \or luglio \or
agosto \or settembre \or ottobre \or novembre \or dicembre
\fi/\number\year;\ \ora}
\else
\def\data{\Di}
\fi

\def\DATA{\number\day\ \ifcase\month\or{}gennaio{}\or%
febbraio{}\or{}marzo{}\or{}aprile\or{}maggio{}\or{}giugno{}\or{}luglio{}\or%
agosto{}\or{}settembre{}\or{}ottobre{}\or{}novembre{}\or{}dicembre\fi%
\ \number\year}

\setbox200\hbox{$\scriptscriptstyle \data $}
\newcount\pgn \pgn=1
\def\foglio{\number\numsec:\number\pgn
\global\advance\pgn by 1} \def\foglioa{A\number\numsec:\number\pgn
\global\advance\pgn by 1}
\global\newcount\numsec\global\newcount\numfor \global\newcount\numfig
\gdef\profonditastruttura{\dp\strutbox}
\def\senondefinito#1{\expandafter\ifx\csname#1\endcsname\relax}
\def\SIA #1,#2,#3 {\senondefinito{#1#2} \expandafter\xdef\csname
#1#2\endcsname{#3} \else \write16{???? ma #1,#2 e' gia' stato definito
!!!!} \fi} \def\etichetta(#1){(\veroparagrafo.\veraformula) \SIA
e,#1,(\veroparagrafo.\veraformula) \global\advance\numfor by 1
\9{\write15{\string\FU (#1){\equ(#1)}}} \9{ \write16{ EQ \equ(#1) == #1
}}} \def \FU(#1)#2{\SIA fu,#1,#2 }
\def\etichettaa(#1){(A\veroparagrafo.\veraformula) \SIA
e,#1,(A\veroparagrafo.\veraformula) \global\advance\numfor by 1
\9{\write15{\string\FU (#1){\equ(#1)}}} \9{ \write16{ EQ \equ(#1) == #1
}}} \def\getichetta(#1){Fig.  \verafigura \SIA e,#1,{\verafigura}
\global\advance\numfig by 1 \9{\write15{\string\FU (#1){\equ(#1)}}} \9{
\write16{ Fig.  \equ(#1) ha simbolo #1 }}} \newdimen\gwidth \def\BOZZA{
\def\alato(##1){ {\vtop to \profonditastruttura{\baselineskip
\profonditastruttura\vss
\rlap{\kern-\hsize\kern-1.2truecm{$\scriptstyle##1$}}}}}
\def\galato(##1){ \gwidth=\hsize \divide\gwidth by 2 {\vtop to
\profonditastruttura{\baselineskip \profonditastruttura\vss
\rlap{\kern-\gwidth\kern-1.2truecm{$\scriptstyle##1$}}}}} }
\def\alato(#1){} \def\galato(#1){}
\def\veroparagrafo{\number\numsec}\def\veraformula{\number\numfor}
\def\verafigura{\number\numfig}
\def\geq(#1){\getichetta(#1)\galato(#1)}
\def\Eq(#1){\eqno{\etichetta(#1)\alato(#1)}}
\def\eq(#1){\etichetta(#1)\alato(#1)}
\def\Eqa(#1){\eqno{\etichettaa(#1)\alato(#1)}}
\def\eqa(#1){\etichettaa(#1)\alato(#1)}
\def\eqv(#1){\senondefinito{fu#1}$\clubsuit$#1\write16{No translation
for #1} \else\csname fu#1\endcsname\fi}
\def\equ(#1){\senondefinito{e#1}\eqv(#1)\else\csname e#1\endcsname\fi}
\openin13=#1.aux \ifeof13 \relax \else \input #1.aux \closein13\fi
\openin14=\jobname.aux \ifeof14 \relax \else \input \jobname.aux
\closein14 \fi \9{\openout15=\jobname.aux} \newskip\ttglue


\font\ottorm=cmr8\font\ottoi=cmmi7\font\ottosy=cmsy7
\font\ottobf=cmbx7\font\ottott=cmtt8\font\ottosl=cmsl8\font\ottoit=cmti7
\font\sixrm=cmr6\font\sixbf=cmbx7\font\sixi=cmmi7\font\sixsy=cmsy7

\font\fiverm=cmr5\font\fivesy=cmsy5\font\fivei=cmmi5\font\fivebf=cmbx5
\def\ottopunti{\def\rm{\fam0\ottorm}\textfont0=\ottorm%
\scriptfont0=\sixrm\scriptscriptfont0=\fiverm\textfont1=\ottoi%
\scriptfont1=\sixi\scriptscriptfont1=\fivei\textfont2=\ottosy%
\scriptfont2=\sixsy\scriptscriptfont2=\fivesy\textfont3=\tenex%
\scriptfont3=\tenex\scriptscriptfont3=\tenex\textfont\itfam=\ottoit%
\def\it{\fam\itfam\ottoit}\textfont\slfam=\ottosl%
\def\sl{\fam\slfam\ottosl}\textfont\ttfam=\ottott%
\def\tt{\fam\ttfam\ottott}\textfont\bffam=\ottobf%
\scriptfont\bffam=\sixbf\scriptscriptfont\bffam=\fivebf%
\def\bf{\fam\bffam\ottobf}\tt\ttglue=.5em plus.25em minus.15em%
\setbox\strutbox=\hbox{\vrule height7pt depth2pt width0pt}%
\normalbaselineskip=9pt\let\sc=\sixrm\normalbaselines\rm}

\catcode`@=11
\def\footnote#1{\edef\@sf{\spacefactor\the\spacefactor}#1\@sf
\insert\footins\bgroup\ottopunti\interlinepenalty100\let\par=\endgraf
\leftskip=0pt \rightskip=0pt \splittopskip=10pt plus 1pt minus 1pt
\floatingpenalty=20000
\smallskip\item{#1}\bgroup\strut\aftergroup\@foot\let\next}
\skip\footins=12pt plus 2pt minus 4pt\dimen\footins=30pc\catcode`@=12
\let\nota=\ottopunti

\newdimen\xshift \newdimen\xwidth \newdimen\yshift
\def\ins#1#2#3{\vbox to0pt{\kern-#2pt \hbox{\kern#1pt #3}\vss}\nointerlineskip} 

\def\eqfig#1#2#3#4#5{ \par\xwidth=#1pt
\xshift=\hsize \advance\xshift by-\xwidth \divide\xshift by 2
\yshift=#2pt \divide\yshift by 2 \line{\hglue\xshift \vbox to #2pt{\vfil #3
\includegraphics{#4} }\hfill\raise\yshift\hbox{#5}}} 

\newwrite\figura   \def\8{\immediate\write\figura\bgroup}


\def\V#1{{\,\underline#1\,}}
\def\T#1{#1\kern-4pt\lower9pt\hbox{$\widetilde{}$}\kern4pt{}}
\let\dpr=\partial \let\io=\infty\let\ig=\int
\def\fra#1#2{{#1\over#2}}\def\media#1{\langle{#1}\rangle}\let\0=\noindent
\def\guida{\leaders\hbox to 1em{\hss.\hss}\hfill}
\def\tende#1{\vtop{\ialign{##\crcr\rightarrowfill\crcr
\noalign{\kern-1pt\nointerlineskip} \hglue3.pt${\scriptstyle
#1}$\hglue3.pt\crcr}}} \def\otto{\
{\kern-1.truept\leftarrow\kern-5.truept\to\kern-1.truept}\ }

\def\tto{{\Rightarrow}}
\def\pagina{\vfill\eject}

\def\*{\vskip0.3truecm}

\def\lis#1{{\overline #1}}\def\eg{\hbox{\it e.g.\ }}

\def\ie{\hbox{\it i.e.\ }}

\def\fiat{{}}
\def\\{\hfill\break} \def\={{ \; \equiv \; }}

\def\annota#1{\footnote{${{}^{{\bf#1\relax \rm}}}$}}
\ifnum\aux=1\BOZZA\else\relax\fi
\ifnum\tipoformule=1\let\Eq=\eqno\def\eq{}\let\Eqa=\eqno\def\eqa{}
\def\equ{{}}\fi
\def\defi{\,{\buildrel def \over =}\,}

\def\1{\ifnum\mgnf=0\pagina\else\relax\fi}
\def\W#1{#1_{\kern-3pt\lower6.6truept\hbox to 1.1truemm
{$\widetilde{}$\hfill}}\kern2pt\,}

\def\xx{{\V x}}

\def\NN{{\cal N}}\def\CC{{\cal C}}\def\EE{{\cal E}}

\def\aa{{\V \a}}

\def\ndpr{{\kern1pt\raise 1pt\hbox{$\not$}\kern1pt\dpr\kern1pt}}
\def\Ndpr{{\kern1pt\raise 1pt\hbox{$\not$}\kern0.3pt\dpr\kern1pt}}

\def\DATA{\number\day\ \ifcase\month\or{}gennaio{}\or%
febbraio{}\or{}marzo{}\or{}aprile\or{}maggio{}\or{}giugno{}\or{}luglio{}\or%
agosto{}\or{}settembre{}\or{}ottobre{}\or{}novembre{}\or{}dicembre\fi%
\ \number\year}
\def\DATA{1 agosto 1998}
\def\qq{{\V q}}\def\yy{{\V y}}\def\PP{{\cal P}}\def\LL{{\cal L}}
\fiat


\def\bf{\caps}

\0{\bf \S1. Chaotic motions.}\annota{*}{\rm Expanded text of the talk at the
ICM98 in Berlin, 26 August 1998.}
\numsec=1\numfor=1\*

\noindent
A typical system exhibiting chaotic motions is a gas in a box whose
particles interact via short range forces with a repulsive core, \eg
a hard core. No hope to ever be able to solve the evolution equations.

In the very simple case of pure hard cores it has been possible to
prove, mathematically at least in some cases, that the system is
ergodic, [Si], [Sz], but ergodicity in itself is only a beginning of
the qualitative theory of the motion. A similar situation arises in
Fluid Mechanics: is a qualitative theory of Turbulence possible as,
clearly, there are hopes to be able, in the near future, to prove
an existence--uniqueness theorem but there is no hope for exact
solutions of Navier Stokes equations?

Equilibrium Statistical Mechanics is a brilliant example of a very
successful quantitative theory derived from a comprehensive qualitative
hypothesis, the {\it ergodic hypothesis}. The key to its success is a
{\it general} expression for the probability distribution
$\m$ on phase space $M$ providing us with the {\it statistics} of
the motions corresponding to given values of the macroscopic
parameters determining the state of the system.

The statistics $\m$ is defined in terms of the time
evolution map $S$ via the relation:

$$\lim_{T\to\io}\fra1T \sum_{j=0}^{T-1} F(S^jx)=\ig_M
F(y)\m(dy)\Eq(1.1)$$
for all smooth {\it observables} $F$ and for almost all, in the sense
of volume measure on $M$, initial data $x\in M$. In Equilibrium
Statistical Mechanics the distribution $\m$ is identified with the
uniform distribution on the surface of constant energy (the
macroscopic state of the system being detemined by the volume $V$ of the
container box and by the energy $U$), which is an obviously invariant
distribution by Liouville's theorem of Hamiltonian Mechanics: this is
a necessary consequence of the ergodic hypothesis.

The success of Equilibrium Statistical Mechanics can be traced back to
the fact that the ergodic hypothesis provides us with a concrete
general, {\it model independent}, expression for the statistics of the
motions. An expression that can be used to derive relations among {\it
time averages} of various observables without even dreaming of ever
being able to actually compute any of such averages.

The Boltzmann's {\it heat theorem}, the positivity of compressibility
and specific heat are simple, but great, examples of such
relations. They are relations which hold for any model, provided one
makes the ergodicity hypothesis, see [Ga1]. A classical argument that
can be used to derive the heat theorem (\ie the second law of
Thermodynamics) from ergodicity is provided us by Boltzmann, see
Appendix A2 and [Ga2].

Consider a mechanical system: viewing its phase space as a discrete
set of points the ergodic hypothesis says that motion is a one cycle
permutation of the points. Given a initial datum with energy $U$ and
with volume $V$ we define {\it temperature} the time average of
kinetic energy $T=\media{K}$ and {\it pressure} the time average of
the derivative of the potential $\f$ with respect to the volume $V$
(note that the force acting on the particles consists of the internal
pair forces {\it and} of the force that the walls exercize upon the
particles which depends on the position of the walls, hence it does
change when the volume varies). Here and below $\media{F}$ will denote
the time average of the observable $F$.

A general elementary property of a system whose motion on each energy
surface is a single periodic motion is that if one calls
$p=\media{\dpr_V \f}$ then:

$$\fra{dU+p \, dV}T=\ exact\Eq(1.2)$$
which means that if the energy $U$ and a parameter $V$ on which the
potential depends (it will be the volume in our case) are varied by
$dU$ and $dV$ respectively {\it then the differential in \equ(1.2) is
exact}.

An elementary classical calculation shows that $p$, see Appendix A2,
in the case of a gas in a box, has the meaning of average force
exercized per unit surface on the walls of the container as a
consequence of the particles collisions: thus we see that the ergodic
hypothesis plus a general, trivial, identity among the averages of
suitable mechanical quantities yields a relation (``equality of cross
derivatives'') holding without free parameters.

The reason why such relation is physically relevant for macroscopic
systems is that the time necessary for the averages defining $T,p$ to
be reached within a good approximation by the finite time averages of
$K, \dpr_V\f$ is not the unobservable {\it recurrence time} (\ie the
superastronomic time for the system to complete a single tour of the
energy surface $U$) but it is a much shorter physically observable time
(whose theory is also due to Boltzmann being the essence of the
Boltzmann's equation) because the quantities $K,\dpr_V\f$ have an
essentially constant value on the energy surface if the number of
particles is large (so that the average of such observables
``stabilizes'' very rapidly compared to the recurrence times).

To summarize: a simple hypothesis allows us to find the statistics of
the motions of an equilibrium system: this implies simple parmeterless
relations among averages of physically relevant quantities (\ie
$\dpr_V\fra1T=\dpr_U\fra{p}T$) which are observable in large systems
because such quantities average very quickly compared to the recurrence
times (being practically constant on the surface of given energy if
the system is large).

Thus a natural question arises: is there anything analogous in Non
Equilibrium Statistical Mechanics? and in developed Turbulence?

The first problem is ``what is the analogue of the uniform Liouville's
distribution?''. This is a really non trivial question that, once
answered, will possibly allow us to try to find relations between time
averages of mechanical quantities. The nontriviality is due to the
fact that as soon as a system is out of equilibrium, \ie
nonconservative forces act upon it, dissipation is necessary in order
to be able to reach a stationary state. But this means that {\it any}
model used will be necessarily described by an evolution equation
which will have a nonzero divergence: so that phase space will
necessarily contract, in the average, and the statistics of the motion
will be concentrated on a set of zero Lebesgue volume, see [Ru3].

Ruelle's proposal in the early 1970's was that one should regard such
systems as {\it hyperbolic} so that there would be a unique stationary
distribution describing the statistics of almost all initial data
(chosen with the uniform distribution on phase space), [Ru1].

This principle has been interpreted in [GC] as the following:
\*
\0{\it Chaotic hypothesis: A chaotic mechanical system can be regarded
for practical purposes as a topologically mixing Anosov system.}
\*

This means that the closure of the attractor is a smooth surface on
which the evolution is a Anosov system: of course assuming Axiom A
instead of Anosov would be more natural, particularly in few degrees
of freedom systems, [Ru1]. However I prefer to formulate the
hypothesis in terms of Anosov system as fractality of the closure of
the attractor seems to be of little relevance in systems with large
number of degrees of freedom occurring in Statistical Mechanics.

The locution {\it practical purposes} is deliberately ambiguous as we
know that even in Equilibrium Statistical Mechanics the corresponding
ergodic hypothesis may fail while its consequences, at least some of
them, will not (like the heat theorem in a free gas or in a harmonic
chain).

The above physical discussion serves as a quick motivation of the
mathematical question: {\it are there general properties shared by
mechanical systems that are transitive or mixing Anosov systems?}.

In the next sections I provide some affirmative answer in the class
of {\it time reversible Anosov maps} and of {\it weakly interacting
chains of Anosov maps}. Recall: a {\it time reversal} symmetry
for a dynamical system $(M,S)$ is {\it any} volume preserving
diffeomorphism $I$ such that:

$$I^2=1,\qquad I\,S= S^{-1}\, I\Eq(1.3)$$
Examples in Hamiltonian mechanical systems are the velocity reversal,
or the composition of the velocity reversal and the parity symmetry,
or the composition of the velocity reversal, parity symmetry and
charge conjugation symmetry. In general a time reversal may be a
symmetry quite different from the naive one that can be imagined, see
[BG].

Hamiltonian systems on which further anholonomic constraints are
imposed via Gauss' principle of {\it least constraint} often generate
systems which show a time reversal symmetry, see Appendix A1, thus
providing the simplest examples.
\*

\0{\bf\S2. Time reversible dissipative Anosov systems. Fluctuation theorem.}
\numsec=2\numfor=1\*

\noindent
We now study a $C^\io$, topologically mixing, Anosov system $(M,S)$ on
a compact manifold $M$.

Let $ M$ be a $d$--dimensional, $C^\io$, compact manifold and let $S$
be a $C^\io$, mixing (transitive would suffice) Anosov diffeomorphism,
[AA], [Si]. If $W^u_x,W^s_x$ denote the {\it unstable} or {\it stable}
manifold at $x\in M$, we call $W^{u,\d}_x,W^{s,\d}_x$ the connected
parts of $W^u_x,W^s_x$ containing $x$ and contained in the sphere with
center $x$ and radius $\d$. Let $d_u,d_s$ be the {\it dimensions} of
$W^u_x,W^s_x$: $d=d_u+d_s$. We shall take $\d$ always smaller than the
smallest curvature radius of $W^u_x,W^s_x$ for $x\in M$. Transitivity
implies that $W^u_x,W^s_x$ are dense in $ M$ for all $x\in M$.

The map $S$ can be regarded, locally near $x$, either as a map of $ M$
to $ M$ or of $W^u_x$ to $W^u_{Sx}$, or of $W^s_x$ to $W^s_{Sx}$.  The
{\it Jacobian matrices} of the "three" maps will be $d\times d$,
$d_u\times d_u$ and $d_s\times d_s$ matrices denoted respectively
$\dpr S(x), \,\dpr S(x)_u,\,\dpr S(x)_s$.  The absolute values of the
respective determinants will be denoted $\L(x)$, $\L_u(x)$, $\L_s(x)$
and are H\"older continuous functions, strictly positive (in fact
$\L(x)$ is $C^\io$), [Si], [AA], [Ru4]. Likewise one can define the
Jacobians of the $n$--th iterate of $S$; they are denoted by appending
a label $n$ to $\L,\L_u,\L_s$ and are related to the latter by the
differentiation chain rule:

$$\eqalign{\L_n(x)=&\prod_{j=0}^{n-1}\L(S^jx),\quad\L_{u,n}(x)=
\prod_{j=0}^{n-1}\L_u(S^jx),\cr
\quad\L_{s,n}(S^jx)=&\prod_{j=0}^{n-1}\L_{s,n}(S^jx),\qquad
\L_n(x)=\L_{u,n}(x)\,\L_{s,n}(x)\chi_n(x)\cr}\Eq(2.1)$$

\0and $\chi_n(x)=\fra{\sin\a(S^nx)}{\sin\a(x)}$ is the ratio of the
sines of the {\it angles} $\a(S^nx)$ and $\a(x)$ between $W^u$ and $W^s$ at
the points $S^nx$ and $x$.  Hence $\chi_n(x)$ is bounded above and
below in terms of a constant $B>0$: $B^{-1}\le\chi_n(x)\le B$, for all
$x$ (by the transversality of $W^u$ and $W^s$).

We can define the {\it forward} and {\it backward statistics} or {\it
``{\rm SRB} distributions''} $\m_+,\m_-$ of the volume measure $\m_0$ via
the limits:

$$\lim_{T\to\io}\fra1T\sum_{k=0}^{T-1} F(S^{\pm k}x)=\ig_\CC \m_\pm(dy)
F(y)\=\m_\pm (F)\Eq(2.2)$$
which {\it exist for all smooth functions $F$ on $M$ and for all but a set
of zero volume of initial points $x$}, see [Si].

Therefore it is the probability distribution $\m_+$ that is the {\it
statistics} $\m$ of the motions (almost surely with respect to the
volume measure $\m_0$ on $M$), see \equ(1.1): {\it it plays the role of the
Gibbs distribution, or microcanoncial ensemble, of equilibrium
Statistical Mechanics}. Hence we are looking for general properties
of $\m_+$, independent of the system considered, if possible.

Let $\L(x)=|\det\dpr S(x)|$; let $\m_\pm$ be the forward and backward
statistics of the volume measure $\m_0$ (\ie the SRB distributions for
$S$ and $S^{\,-1}$).\*

\0{\it Definition: The system $( M ,S)$ is {\sl dissipative} if:

$$-\ig_{ M }\m_\pm(dx)\log\L^{\,\pm1}(x)=\lis\h_\pm>0\Eq(2.3)$$}

\0{\it Remarks:}
1) Existence of a time reversal symmetry $I$, see \equ(1.3), implies
$\lis\h_+=\lis\h_-$ and $\L(x)=\L^{\,-1}(I\,x)$; furthermore
$I\,W^u_x=W^s_{I\,x}$ and the dimensions of the stable and unstable
manifolds $d_s,d_u$ are equal: $d_u=d_s$ and $d=d_u+d_s$ is even.
\\
2) if $\L_u(x),\L_s(x)$ denote the absolute values of the Jacobian
determinants of $S$ as a map of $W^u_x$ to $W^u_{Sx}$ and of
$W^s_x$ to $W^s_{Sx}$, then $\L_u(x)=\L_s(I\,x)^{\,-1}$.
\\
3) If a system $( M ,S)$ is dissipative then the system $( M ',S')$
with $ M '= M \times M $ and $S'(x,y)=(Sx, S^{-1}y)$ provudes us with
an example, setting $I(x,y)=(y,x)$, of a dynamical system in the
general class of ``reversible'' Anosov maps considered in \S1.
\*
{\it From now on only reversible dissipative Anosov dynamical systems
$( M ,S)$ will be considered}: it is for such systems that it will be
possible to derive general model independent properties.
\*
\0{\it Definition: The {\sl ``dimensionless entropy production rate''} or
the {\sl ``phase space contraction rate''} at $x\in M $ and
over a time $\t$ is the function $\e_\t(x)$:

$$x\to\e_\t(x)=\fra1{\lis\h_+\t}\sum_{j=-\t/2}^{\t/2-1}\log
\L^{\,-1}(S^jx)=\fra1{\lis\h_+\t}\log \lis\L^{\,-1}_\t(x)\Eq(2.4)$$

\0with $\lis\L_\t(x){\buildrel def \over =}\prod_{-\t/2}^{\t/2-1}
\L(S^jx)$. Hence  (see \equ(2.2)) it is, with $\m_0$--probability $1$:

$$\media{\e_\t}_+=\lim_{T\to+\io}
\fra1T\sum_{j=0}^{T-1}\e_\t(S^jx) \=\ig_ M \m_+(dy)\e_\t(y)=1\Eq(2.5)$$
}

{}From the general theory of Anosov systems, [Si], it follows that the
$\m_+$--probability that $p=\e_\t(x)$ is in the interval $[p-\d,p+\d]$
can be written as $\max_{q\in [p-\d,p+\d]} e^{\t\lis\z(q)}$ for some
suitably chosen function $\lis\z(p)$ and up to a factor bounded by
$B^{\pm1}$ with $0<B<+\io$. This is a deep result of Sinai that holds
because the statistics $\m_+$ can be regarded as a Gibbs distribution
and one can use the large deviation theory for such distributions: see
Appendix A3 below for details. Then the following theorem holds, see [GC]:
\*
\0{\it Fluctuation theorem: The {\it ``large deviation function''} $\lis \z(p)$
is analytic in an interval $(-p^*,+p^*)$ with $p^*\ge1$ and verifies
the relation:

$$\fra{\lis\z(p)-\lis\z(-p)}{p\lis\h_+}=1\qquad |p|<p ^*\Eq(2.6)$$
\ie the odd part of $\lis\z(p)$ is in general linear and its slope is
equal to the average entropy creation rate.}
\*

What one really checks, see [Ga3], is the existence of $p^*\ge1$ such
that the SRB distribution $\m_+$ verifies:

$$p-\d\le
\lim_{\t\to\io}\fra1{\lis\h_+\t}\log\fra{\m_+(\{\e_\t(x)\in[p-\d,p
+\d]\})}{\m_+(\{\e_\t(x)\in-[p-\d,p+\d]\})}\le p+\d\Eq(2.7)$$

\0for all $p, \,|p|<p^*$.
\*
The above theorem was first informally proved in [GC] where its
interest for nonequilibrium statistical mechanics was pointed out.
The theorem can be regarded as a large deviation result for the
probability distribution $\m_+$. Although I think that the physical
interest of the theorem far outweighs its mathematical aspects it is
useful to see a formal proof.  A proof is reproduced in Appendix A3
below: it is taken from [Ga3].

The relation \equ(2.6) has been tested numerically in several cases:
it was in fact discovered in a numerical experiment, see [ECM2], and
tested in other experiments, see [BGG], [BCL], [LLP]. Why does one
need to test a theorem? the reason is that in concrete cases not only
it is not known whether the system is Anosov but, in fact, it is
usually clear that it is not, see [RT]. Hence the test is necessary to
check the Chaotic Hypothesis which says that the failure of the Aoosov
property should be irrelevant for ``practical purposes''.

Another interesting aspect, that cannot be treated here for
limitations of time, of the above theorem is that it can be
interpreted as an extension to non zero forcing (\ie $\lis\h_+>0$) of
the Green--Kubo relations: see [Ga6].
\*
\0{\bf \S3. Fluctuations in large systems.}
\numsec=3\numfor=1\*

\noindent
An important drawback of the above fluctuation theorem, besides the
reversibility assumption which is not verified in many important
cases, is that it can be practically verified, for physical as well as
mathematical reasons, only in (relatively) small systems.

In fact the logarithm of the entropy creation rate distribution $\t\,
\lis\z(p)$ is, usually, not only proportional to $\t$, \ie to the time
interval over which the entropy creation fluctuation is observed, but
{\it also to the spatial extension of the system}, \ie to the number of
degrees of freedom; so that it is extremely unlikely that observing
$p$ in a large system one can see a value $p$ which is appreciably
different from $1$ (note that the normalizing constant $\lis\h_+$ in
\equ(2.4) is so chosen that the average of $p$ in the stationary state
is $1$).

For this reason in macroscopic (or just ``large'') systems the phase
space contraction rate is essentially constant (and its physical
interpretation is of strength of the friction) much as the density is
constant in gases at equilibrium. Therefore one can hope to see
entropy creation rate f\/luctuations only if one can define a {\it
local entropy creation rate} $\h_{V_0}(x)$ associated with a
microscopic region $V_0$ of space.

I now discuss, heuristically, why one should expect that a {\it local
entropy creation rate can be defined, at least in some cases, and
verifies a local version of the f\/luctuation law} \equ(2.6).  This is
discussed in a special example, see [Ga7], as in general one can doubt that a
local version of the fluctuation law holds, see [BCL].

The special example that we select is the {\it chain} of weakly
coupled Anosov maps, well studied in the literature, [PS]. The system
has a translation invariant spatial structure, \ie it is a chain (or a
lattice) of weakly interacting chaotic (mixing Anosov) system. This
can be described as follows.

Let $(M',S')$ be a dynamical system whose phase space $M'$ is a product
of $2N+1$ identical analytic manifolds $\lis M_0$: $M'=\lis M_0^{2N+1}$
and $S': \,M' \to M'$ is a small perturbation of a product map $\lis
S_0\times \ldots\times\lis S_0\defi \tilde S_0$ on $M'$. We assume that
$(\lis M_0,\lis S_0)$ is a mixing Anosov systems. The size $N$ (an
integer) will be called the ``spatial size'' of the system.

For $x,y,z\in \lis M_0$ let $F_\e(x,z,y)$ be analytic and such that
$z\to F_\e(x,z,y)$ is a map, of $\lis M_0$ into itself, {\it
$\e$--close to the identity} and $\e$--analytic for $|\e|$ small
enough. We suppose that, if $\xx=(x_{-N},\ldots, x_N)\in M'$:
$$(S'\xx)_i= F_\e(x_{i-1},x_i,x_{i+1})\circ S_0 x_i\Eq(3.1)$$
where $x_{\pm(N+1)}$ is {\it identified} with $x_{\mp N}$ (\ie we
regard the chain as periodic); we call such a dynamical system a {\it
chain of interacting Anosov maps} coupled by nearest neighbors. It is
a special example of the class of maps considered in [PS].\annota1{\rm
In the paper [PS] it is assumed that {\it also} $\lis S_0$ (hence
$S_0$) is close to the identity, \eg within $\e$: such condition does
not seem necessary for the purposes of the present paper, hence it
will not be assumed.}

It is difficult, maybe even impossible, to construct a (non trivial)
reversible system of the above form: we therefore (see [Ga3]) consider
the system $(M,S)$ where $M=M'\times M'$ and define $S_0\defi\tilde
S_0\times (\tilde S_0)^{-1}$ and $S\defi S'\times (S')^{-1}$, called
hereafter the {\it free evolution} and the {\it interacting
evolution}, respectively. So that the system can be considered as time
reversible with a time reversal map $I(\xx,\yy)=(\yy,\xx)$.  Note that
the inverse map to \equ(3.1) does not have the same form. The map $S$
is, however, still in the class considered in [PS] because it can be
written as $S(\xx,\yy)_i=\big(S(\xx,\yy)_{i1},S(\xx,\yy)_{i2}\big)$
with:
$$\eqalign{
S(\xx,\yy)_{i1}=& F_\e(x_{i-1},x_i,x_{i+1})\circ S_0\, x_i\cr
S(\xx,\yy)_{i2}=& G_{\e,i}(\yy)\circ S_0^{-1} y_i\cr
}\Eq(3.2)$$
where $G$ has ``short range'', \ie $|G_\e(\yy)_i-G_\e(\yy')_i|$ is of
order $\e^k$ if $\yy$ and $\yy'$ coincide on the sites $j$ with
$|j-i|\le k$. By definition the system $(M,S)$ is ``{\it
reversible}'', \ie the volume preserving diffeomorphism $I$ verifies
\equ(1.3) above.

Therefore the points of the phase space $M$ will be $(\xx,\yy)=
(x_{-N},y_{-N},\ldots, x_N,y_N)$: however, to simplify notations, we
shall denote them by $\xx=(x_{-N},\ldots,x_N)$, with $x_j$ denoting,
of course, a {\it pair} of points in $\lis M_0$.

If $\e$ is small enough the interacting system will still be
hyperbolic, \ie for every point $\xx$ it will be possible to define a
stable and an unstable manifolds $W^s_\xx,W^u_\xx$, [PS], so that the
key notion of ``Markov partition'', [Si], will make
sense and it will allow us to transform, following the work [PS], the
problem of studying the statistical properties of the dynamics of the
system into an equivalent, but much more familiar, problem in
equilibrium statistical mechanics of lattice spin systems interacting
with short range forces. The reader will recognize below that this
method is the natural extension to chains of the method used in
Appendix A3 to study a single Anosov system.

The main notion that we want to introduce for our chain is the
notion of {\it local entropy creation rate} $\h_{V_0}(\xx)$, the
entropy creation rate inside a fixed finite set $V_0\subset [-N,N]$ of
Anosov systems among the $2N+1$ composing the chain.
\*
\0{\it Definition:
Fixed a point $\xx=(\ldots, x_{\ell-1}, x_{\ell}, x_{\ell+1},\ldots)$
consider the map \equ(3.1) as a map of
$\xx_{V_0}\defi(x_j)_{j\in V_0}=(x_{-\ell},\ldots,x_\ell)$ into:
$$\xx'_{V_0}=S(\ldots,x_{-\ell-1},\xx_{V_0},x_{\ell+1},\ldots)_{V_0}
\Eq(3.3)$$
defined by \equ(3.1) for $i\in[-\ell,\ell]$. We call {\it ``local
entropy production rate''} associated with the ``space like box''
$V_0=[-\ell,\ell]$ at the phase space point
$\xx=(\ldots,x_{\ell-1},x_{\ell},x_{\ell+1},\ldots)$ the quantity
$\h_{V_0}(\xx)$ equal to {\sl minus the logarithm of the determinant
of the $2(2\ell+1)\times2(2\ell+1)$ Jacobian matrix of the map}
\equ(3.3).}
\*

Given a finite region $V_0$ centered at the origin and a time
interval $T_0$, let $\h_+$ denote the average density of entropy
creation rate, \ie $\h_+=\lim_{V_0, T_0\to\io}
\fra1{|T_0|}\fra1{|V_0|}\sum_{j=0}^{|T_0|-1}\h_{V_0}(S^j x)$, then we set:
$$p=\fra1{\h_+ |V|} \sum_{j=-\fra12|T_0|}^{\fra12 |T_0|} \h_{V_0}(S^j
x), \qquad V=V_0\times T_0\Eq(3.4)$$
where $\h_{V_0}(x)$ denotes the entropy creation rate in the region
$V_0$.

Calling $\p_V(p)$ the probability distribution of $p$ in the
stationary state $\m_+$, \ie in the SRB distribution, {\it and assuming
that the system is a weakly coupled chain of Anosov systems} I shall
show, heristically, that:
\*
\0{\it Proposition: It is $\p_V(p)= e^{\z(p)|V|+O(|\dpr V|)}$ where
$|\dpr V|$ denotes the size of the boundary of the space--time region
$V$ and $\z(p)$ is a function analyticin $p\in (-p^*,p^*)$ for some
$p^*\ge1$.  And:
$$\eqalign{
&\fra{\z(p)-\z(-p)}{p\,\h_+}=1, \qquad |p|<p^*\cr
&\lis\z(p)=r\, \z(p),\qquad\lis\h_+=r\, \h_+\cr}\Eq(3.5)$$
where $r$ is the ratio between the total ``volume'' ($2N+1$) of the
system and the volume $V_0$, \ie the ``global'' and ``local'' distributions
are trivially related if appropriately normalized.}
\*

Note that this implies that if $V_0$ is an interval of length
$L=|V_0|$ and if $H=|T_0|$ then the relative size of the error and of
the leading term will be, for some length $R$, of order $(L+H)R$
compared to order $LH$. Hence a relative error $O(H^{-1}+L^{-1})$ is
made by using simply $\z(p)$ to evaluate the logarithm of the
probability of $p$ as defined by \equ(3.4)).

The interest of the above statements lies in their independence on the
total size $2N+1$ of the systems and the relevance of the above
proposition for concrete applications should be clear.

{\it It means that the fluctuation theorem leads to observable
consequences if one looks at the far more probable microscopic
f\/luctuations of the local entropy creation rate}. One can test the
relation \equ(3.5) in a small region $V_0$ even when the system is
very large: in such regions the entropy creation rate fluctuations
will be frequent enough to be observable and carefully
measurable. These fluctuations behave, therefore, just as ordinary
density fluctuations at equilibrium: also the latter are not
macroscopically observable but they are easily observable in small
volumes.

The key results for the analysis leading to the above proposition are
the papers [GC], [Ga3] and, mainly, [PS]: the latter paper provides us
with a deep analysis of chains of Anosov systems and it contains, I
believe, all the ingredients necessary to make the analysis
mathematically rigorous: however I do not attempt at a mathematical
proof here. The analysis is presented in Appendix A4 below.

Other types of fluctuation theorems (concerning non SRB distributions)
had been previously found, see [ES]; extensions to stochastic systems
have been recently discussed, see [Ku], [LS].

\*
\0{\bf Acknowledgments:\it\ I have profited of stimulating discussions
with F. Perroni, who also helped with numerical tests of the
above ideas, with F. Bonetto and D. Ruelle. This work is part of the
research program of the European Network on: ``Stability and
Universality in Classical Mechanics", \# ERBCHRXCT940460; partially
supported also by Rutgers University and CNR-GNFM.}
\ifnum\mgnf=0\pagina\fi

\*
\0{\bf Appendix A1:\it\ The Gauss' minimal constraint principle.}
\numsec=1\numfor=1\pgn=1 \*

\noindent
Let $\f(\dot \xx,\xx)=0$, $\xx=\{\dot\xx_j,\xx_j\}$ be a constraint
and let $\V R(\dot\xx,\xx)$ be the constraint reaction and $\V
F(\dot\xx,\xx)$ the active force, see also Appendix A1 of [Ga3].

Consider all the possible accelerations $\V a$ compatible with the
constraints and a given initial state $\dot\xx,\xx$. Then $\V R$ is
{\it ideal} or {\it verifies the principle of minimal constraint} if the
actual accelerations $\V a_i=\fra1{m_i} (\V F_i+\V R_i)$ minimize the
{\it effort}:

$$\sum_{i=1}^N\fra1{m_i} (\V F_i-m_i\V a_i)^2\ \otto\
\sum_{i=1}^N (\V F_i-m_i\V a_i)\cdot\d \V a_i=0\Eqa(A1.1)$$

\0for all possible variations $\d \V a_i$ compatible with the
constraint $\f$. Since all possible accelerations following
$\dot\xx,\xx$ are such that $\sum_{i=1}^N
\dpr_{\dot\xx_i}\f(\dot\xx,\xx)\cdot\d \V a_i=0$ we can write:

$$\V F_i-m_i\V a_i-\a\,\dpr_{\dot\xx_i} \f(\dot\xx,\xx)=\V0\Eqa(A1.2)$$

\0with $\a$ such that $\fra{d}{dt}\f(\dot\xx,\xx)=0$, \ie:

$$\a=\fra{\sum_i\,(\dot\xx_i\cdot\dpr_{\xx_i} \f+\fra1{m_i} \V
F_i\cdot\dpr_{\dot\xx_i}\f)}{\sum_i m_i^{-1}(\dpr_{\dot\xx_i}\f)^2}
\Eqa(A1.3)$$

\0which is the analytic expression of the Gauss' principle, see [LA].

Note that if the constraint is even in the $\dot\xx_i$ then $\a$ is
odd in the velocities: therefore if the constraint is imposed on a
system with Hamiltonian $H= K+V$, with $K$ quadratic in the velocities
and $V$ depending only on the positions, and if on the system act
other purely positional forces (conservative or not) then the
resulting equations of motion are reversible if time reversal is
simply defined as velocity reversal.

The gaussian principle has been somewhat overlooked in the Physics
literature in Statistical Mechanics: its importance has been only
recently brought again to the attention, see the review [HHP].  A
notable, though ancient by now, exception is a paper of Gibbs, [Gi],
which develops variational formulas which he relates to the Gauss
principle of least constraint.
\*
\0{\bf Appendix A2. Heat theorem for monocyclic systems.
Evaluation of the average $\media{\dpr_V \f}$.}
\numsec=2\numfor=1\*

\noindent
Consider a $1$--dimensional system with potential $\varphi(x)$ such that
$|\varphi'(x)|>0$ for $|x|>0$, $\varphi''(0)>0$ and $\varphi(x)\,\,
{\vtop{\ialign{#\crcr\rightarrowfill\crcr
\noalign{\kern0pt\nointerlineskip}\hglue3.pt${\scriptstyle
{x\to\infty}}$\hglue3.pt\crcr}}}\,\,+\infty$ (in other words a
$1$--dimensional system in a confining potential). There is only one
motion per energy value (up to a shift of the initial datum along its
trajectory) and all motions are periodic so that the system is {\it
monocyclic}. Assume also that the potential $\varphi(x)$ depends on a
parameter $V$.

One defines {\it state} a motion with given energy $E$ and given
$V$. And:
\*

\halign{#\ $=$\ & #\hfill\cr
$U$ & total energy of the system $\equiv  K+\varphi$\cr
$T$ & time average of the kinetic energy $K$\cr
$V$ & the parameter on which $\varphi$ is suposed to depend\cr
$p$ & $-$ time average of $\partial_V \varphi$\cr}
\*
\noindent{}A state is parameterized by $U,V$ and if such parameters
change by
$dU, dV$ respectively we define:

$$dL=-p dV,\qquad dQ=dU+p dV\Eqa(A2.1)$$
then:
\*
\noindent{}{\sl Theorem} (Helmholtz): {\it the differential
$({dU+pdV})/{T}$ is exact.}
\*
In fact let:

$$S=2\log \int_{x_-(U,V)}^{x_+(U,V)}\sqrt{K(x;U,V)}dx= 2\log
\int_{x_-(U,V)}^{x_+(U,V)}\sqrt{U-\varphi(x)}dx\Eqa(A2.2)$$
($\fra12S$ is the logarithm of the action), so that:

$$S=\fra{\int (dU-\partial_V\varphi(x) dV)
\fra{dx}{\sqrt{K}}}{
\int K\fra{dx}{\sqrt{K}}}\Eqa(A2.3)$$
and, noting that $\fra{dx}{\sqrt K} =\sqrt{\fra2m} dt$, we see that the
time averages are given by integrating with respect to $\fra{dx}{\sqrt
K}$ and dividing by the integral of $\fra{1}{\sqrt K}$. We find
therefore:

$$dS=\fra{dU+p dV}{T}\Eqa(A2.4)$$
Boltzmann saw that this was not a simple coincidence: his interesting
(and healthy) view of the continuum (which he probably never really
considered more than a convenient artifact, useful for computing
quantities describing a discrete world) led him to think that in some
sense {\it monocyclicity was not a strong assumption}.

In general one can call {\it monocyclic} a system with the property
that there is a curve $\ell\to x(\ell)$, parameterized by its
curvilinear abscissa $\ell$, varying in an interval $0< \ell< L(E)$,
closed and such that $ x(\ell)$ covers all the positions compatible
with the given energy $E$.

Let $ x= x(\ell)$ be the parametric equations so that the energy
conservation can be written:

$$\fra12 m\dot\ell^2+\varphi(x(\ell))=E
\Eqa(A2.5)$$
then if we suppose that the potential energy $\varphi$ depends on a
parameter $V$ and if $T$ is the average kinetic energy,
$p=-\langle{\partial_V \varphi}\rangle$ it is, for some $S$:
$$dS=\fra{dE+pdV}{T},\qquad p=-\langle{\partial_V
\varphi}\rangle,\quad T=\langle{K}\rangle\Eqa(A2.6)$$
where $\langle \cdot\rangle$ denotes time average.

The above can be applied to a gas in a box.  Imagine the box
containing the gas to be covered by a piston of section $A$ and
located to the right of the origin at distance $L$: so that $V=AL$.

The microscopic model for the pistion will be a potential
$\lis\f(L-\x)$ if $x=(\x,\h,\z)$ are the coordinates of a
particle. The function $\lis\f(r)$ will vanish for $r>r_0$, for some
$r_0$, and diverge to $+\io$ at $r=0$. Thus $r_0$ is the width of the
layer near the piston where the force of the wall is felt by the
particles that happen to roam there.

Noting that the potential energy due to the walls is $\f=\sum_j
\lis\f(L-\x_j)$ and that $\dpr_V \f=A^{-1}\dpr_L\f$ we must evaluate
the time average of:

$$\dpr_L \f(x)=-\sum_j \lis\f'(L-\x_j)\Eqa(A2.7)$$
As time evolves the particles with $\x_j$ in the layer within $r_0$ of
the wall will feel the force exercized by the wall and bounce
back. Fixing the attention on one particle in the layer we see that it
will contribute to the average of $\dpr_L \f(x)$ the amount:

$$\fra1{\rm total\ time} 2\ig_{t_0}^{t_1}- \lis\f'(L-\x_j)
dt\Eqa(A2.8)$$
if $t_0$ is the first instant when the point $j$ enters the layer and
$t_1$ is the instante when the $\x$--compoent of the velocity vanishes
``against the wall''. Since $-\lis\f'(L-\x_j)$ is the $\x$--component
of the force, the integral is $-2m|\dot\x_j|$ (by Newton's law),
provided $\dot\x_j>0$ of course.

The number of such contributions to the average per unit time are
therefore given by $\r_{wall}\, A\, \ig_{v>0} 2mv\, f(v)\, v\, dv$ if
$\r_{wall}$ is the density (average) of the gas near the wall and
$f(v)$ is the fraction of particles with velocity between $v$ and
$v+dv$. Using the ergodic hypothesis (\ie the microcanonical
ensemble) and the equivalence of the ensembles to evaluate $f(v)$ it
follows that:

$$p\defi \media{\dpr_V\f}= \r_{wall} \b^{-1}\Eqa(A2.9)$$
where $\b^{-1}=k_B T$ with $T$ the absolute temperature and $k_B$ the
Boltmann's constant. That the \equ(A2.9) yields the correct value of
the pressure is well known, see [MP], in Classical Statistical
Mechanics; in fact often it is even taken as microscopic definition of
the pressure.

\*
\0{\bf Appendix A3. A proof of the fluctuation theorem.}
\numsec=3\numfor=1\*

\0{\it(A) Description of the SRB statistics.}
\*
A set $E$ is a {\it rectangle} with {\it center} $x$ and {\it axes}
$\D^u,\D^s$ if:\\
1) $\D^u,\D^s$ are two connected surface elements of $W^u_x,W^s_x$
containing $x$.\\
2) for any choice of $\x\in\D^u$ and $\h\in\D^s$ the local manifolds
$W^{s,\d}_\x$ and $W^{u,\d}_\h$ intersect in one and only one point
$x(\x,\h)=W^{s,\d}_\x\cap W^{u,\d}_\h$. The point $x(\x,\h)$ will also
be denoted $\x\times\h$.\\
3) the boundaries $\dpr\D^u$ and $\dpr\D^s$ (regarding the latter sets
as subsets of $W^u_x$ and $W^s_x$) have zero surface area on $W^u_x$ and
$W^s_x$.\\
4) $E$ is the set of points $\D^u\times\D^s$.

Note that {\it any} $x'\in E$ can be regarded as the center of $E$
because there are $\D^{\prime u},\D^{\prime s}$ both containing $x'$ and
such that $\D^u\times\D^s\= \D^{\prime u}\times\D^{\prime s}$.
Hence each $E$ can be regarded as a rectangle centered at any $x'\in E$
(with suitable axes). See figure.

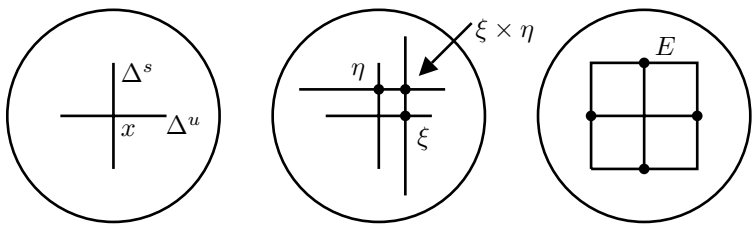
\eqfig{260}{90}{
\ins{43}{37}{$x$}
\ins{43}{60}{$\D^s$}
\ins{60}{40}{$\D^u$}
\ins{155}{36}{$\x$}
\ins{130}{60}{$\h$}
\ins{177}{77}{$\x\times\h$}
\ins{245}{70}{$E$}
}{figurediffxxx.ps}{}
\*
{\nota
\0The circle is a small neighborhood of $x$; the first picture shows the
axes; the intermediate picture shows the $\times$ operation and
$W^{u,\d}_\h, W^{s,\d}_\x$; the third picture shows the rectangle $E$
with the axes and the four marked points are the boundaries $\dpr\D^u$
and $\dpr\D^s$. The picture refers to a two dimensional case and the
stable and unstable manifolds are drawn as flat, \ie the $\D$'s are very
small compared to the curvature of the manifolds. The center $x$ is
drawn in a central position, but it can be {\it any} other point of $E$
provided $\D^u$ and $\D^s$ are correspondingly redefined. One should
meditate on the symbolic nature of the drawing in the cases of higher
dimension.\vfill}

The {\it unstable boundary} of a rectangle $E$ will be the set $\dpr_u
E=\D^u\times\dpr\D^s$; the {\it stable boundary} will be $\dpr_s
E=\dpr\D^u\times\D^s$.  The boundary $\dpr E$ is therefore $\dpr
E=\dpr_s E\cup\dpr_u E$.  The set of the {\it interior points}
of $E$ will be
denoted $E^0$.  A {\it pavement} of $ M$ will be a covering
$\EE=(E_1,\ldots,E_\NN)$ of $ M$ by $\NN$ rectangles with pairwise
disjoint interiors.  The {\it stable (or unstable) boundary} $\dpr_s\EE$
(or $\dpr_u \EE$) of $\EE$
is the union of the stable (or unstable) boundaries of the rectangles
$E_j$: $\dpr_u \EE=\cup_j\dpr_u E_j$ and
$\dpr_s \EE=\cup_j\dpr_s E_j$.

A pavement $\EE$ is called {\it markovian} if its stable boundary
$\dpr_s \EE$ retracts on itself under the action of $S$ and its unstable
boundary retracts on itself under the action of $S^{-1}$, [Si], [Bo], [Ru1];
this means:

$$S\dpr_s\EE\subseteq \dpr_s\EE,\qquad S^{-1}\dpr_u\EE\subseteq \dpr_u\EE
\Eqa(A3.1)$$

\0Setting $M_{j,j'}=0$, $j,j'\in\{1,\ldots,\NN\}$, if $S E^0_j\cap
E^0_{j'}=\emptyset$ and $M_{j,j'}=1$ otherwise we call $C$ the set of
sequences $\V j=(j_k)_{k=-\io}^\io$, $j_k\in\{1,\ldots,\NN\}$ such that
$M_{j_k,j_{k+1}}\=1$.  The transitivity of the system $( M,S)$ implies
that $M$ is {\it transitive}: \ie there is a power of the matrix $M$
with all entries positive.  The space $C$ will be called the space of
the {\it compatible symbolic sequences}.  If $\EE$ is a markovian
pavement and $\d$ is small enough the map:

$$X: \V j\in C\,\to\,x=\bigcap_{k=-\io}^\io S^{-k} E_{j_k}\in M\Eqa(A3.2)$$

\0is continuous and $1-1$ between the complement $ M_0\subset M$
of the set $N=
\cup_{k=-\io}^\io
S^k\dpr \EE $ and the complement $C_0\subset C$ of
$X^{-1}(N)$. This map is called the {\it
symbolic code} of the points of $ M$: it is a code that associates with
each $x\not\in N$ a sequence of symbols $\V j$ which are the labels of
the rectangles of the pavement that are successively visited by the
motion $S^jx$.

The symbolic code $X$ transforms the action of $S$ into the {\it left shift}
$\th$ on $C$: $S X(\V j)= X(\th \V j)$. A key result, [Si], is that it
transforms the {\it volume measure} $\m_0$ on $ M$ into a {\it Gibbs
distribution}, [LR], [Ru2], $\lis\m_0$ on $C$ with formal Hamiltonian:

$$H(\V j)=\sum_{k=-\io}^{-1} h_-(\th^k\V j)+h_0(\V j)+\sum_{k=0}^\io
h_+(\th^k \V j)\Eqa(A3.3)$$

\0where, see \equ(2.1):

$$\eqalign{
h_-(\V j)=&-\log \L_s(X(\V j)),\quad h_+(\V j)=\log \L_u(X(\V j)),\cr
h_0(\V j)=&-\log\sin\a(X(\V j))\cr}\Eqa(A3.4)$$

If $F$ is H\"older continuous on $ M$ the function $\lis F(\V j)=F(X(\V
j))$ can be represented in terms of suitable functions
$\F_k(j_{-k},\ldots,j_k)$ as:

$$\lis F(\V j)=\sum_{k=1}^\io \F_k(j_{-k},\ldots,j_k),\qquad
|\F_k(j_{-k},\ldots,j_k)|\le \f e^{-\l k}\Eqa(A3.5)$$

\0where $\f>0,\l>0$. In particular $h_\pm$ (and $h_0$) enjoy the property
\equ(A3.5) ({\it short range}).

If $\lis\m_+,\lis\m_-$ are the Gibbs states with formal Hamiltonians:

$$\sum_{k=-\io}^\io h_+(\th^k\V j),\qquad
\sum_{k=-\io}^\io h_-(\th^k\V j)\Eqa(A3.6)$$

\0the distributions $\m_\pm$ on $ M$, images of $\lis\m_\pm$ via the
code $X$ in \equ(A3.2), will be the {\it forward} and {\it backward
statistics} of the volume distribution $\m_0$ (corresponding to
$\lis\m_0$ via the code $X$), [Si]. This means that:

$$\lim_{T\to\io}\fra1T\sum_{k=0}^{T-1} F(S^{\pm k}x)=\ig_ M \m_\pm(dy)
F(y)\=\m_\pm (F)\Eqa(A3.7)$$

\0for all smooth $F$ and for $\m_0$--almost all $x\in M$. The distributions
$\m_\pm$ are often called the {\it SRB distributions}, [ER]; the above
statements and \equ(A3.6),\equ(A3.7) constitute the content of a well
known theorem by Sinai, [Si].

An approximation theorem for $\m_+$ can be given in terms of the {\it
coarse graining} of $ M$ generated by the markovian pavement
$\EE_T=\bigvee_{k=-T}^T S^{-k}\EE$.\annota{3}{\nota Where $\vee$
denotes the operation which, given two pavements $\EE,\EE'$ generates a
new pavement $\EE\vee\EE'$: the rectangles of $\EE\vee\EE'$ simply
consist of all the intersections $E\cap E'$ of pairs of rectangles
$E\in\EE$ and $E'\in\EE'$.} If $E_{j_{-T},\ldots,j_T}\=\cap_{k=-T}^T
S^{-k} E_{j_k}$ and $x_{j_{-T},\ldots,j_T}$ is a point chosen in the
coarse grain set $E_{j_{-T},\ldots,j_T}$, so that its symbolic sequence
is obtained by attaching to the right and to the left of
${j_{-T},\ldots,j_T}$ arbitrary compatible sequences depending only on
the symbols $j_{\pm T}$ respectively.  We define the distribution
$\m_{T,\t}$ by setting:

$$\eqalign{\m_{T,\t}(F)\=&\ig_ M \m_{T,\t}(dx) F(x)=
\fra{\sum_{j_{-T},\ldots,j_T}\lis\L_{u,\t}^{\,-1}
(x_{j_{-T},\ldots,j_T}) F(x_{j_{-T},\ldots,j_T})}{\sum_{j_{-T},
\ldots,j_T}\lis\L_{u,\t}^{\,-1}(x_{j_{-T},\ldots,j_T})}\cr
\lis\L_{u,\t}(x){\buildrel def \over =}&
\prod_{k=-\t/2}^{\t/2-1}\L_u(S^kx)\cr}\Eqa(A3.8)$$

Then for all smooth $F$ we have: $\lim_{T\ge\t/2,\,\t\to\io}
\m_{T,\t}(F)=\m_+(F)$. Note that equation \equ(A3.8) can also be
written:

$$\m_{T,\t}(F)=
\fra{\sum_{j_{-T},\ldots,j_T}e^{-\sum_{k=-\t/2}^{\t/2-1} h_+(\th^k\V
j^0)} F(X(\V j^0))} {\sum_{j_{-T}, \ldots,j_T}
e^{-\sum_{k=-\t/2}^{\t/2-1} h_+(\th^k\V j^0)}}\Eqa(A3.9)$$

\0where $\V j^0\in C$  is the compatible sequence agreeing with
$j_{-T},\ldots,j_T$ between $-T$ and $T$ (\ie $X(\V
j^0)=x_{j_{-T},\ldots,j_T}\in E_{j_{-T},\ldots,j_T}$) and continued
outside as above.
\*
\0{\it Notation:} to simplify the notations we shall write, when $T$ is
regarded as having a fixed value, $\qq$ for the elements
$\qq=(j_{-T},\ldots,j_T)$ of $\{1,\ldots,\NN\}^{2T+1}$; and $E_\qq$
will denote $E_{j_{-T},\ldots,j_T}$ and $x_\qq$ the above point of
$E_\qq$.
\*
\0{\it Remark:} Note that the weights in \equ(A3.9) depend on the
special choices of the centers $x_\qq$ (\ie of $\V j^0$); but if $x_\qq$
varies in $E_\qq$ the weight of $x_\qq$ changes by at most a factor,
bounded above by some $B<\io$ and below by $B^{-1}$, for all $T\ge0$,
and essentially depending only on the symbols corresponding to the sites
close to $\pm T$.
\*
The last formula shows that the forward statistics of $\m_0$ can be
regarded as a Gibbs state for a {\it short range one dimensional spin
chain with a hard core interaction}. The spin at $k$ is the value of
$j_k\in\{1,\ldots,\NN\}$; the short range refers to the fact that the
function $h_+(\V j)\=\log \L_u(X(\V j))$, ($\L_u(x)$ {\it being H\"older
continuous}), can be represented as in \equ(A3.5) where the $\F_k$ play
the role of "many spins" interaction potentials and the hard core refers
to the fact that the only spin configurations $\V j$ allowed are those
with $M_{j_k,j_{k+1}}\=1$ for all integers $k$.
\*
\0{\it(B) A Legendre transform.}
\*
First the function \equ(2.4) is converted to a function on the spin
configurations $\V j\in C$:

$$\tilde\e_\t(\V j)=\e_\t(X(\V
j))=\fra1{\t}\sum_{k=-\t/2}^{\t/2-1} L(\th^k\V j)\Eqa(A3.10)$$

\0where $L(\V j)\=\fra1{\lis\h_+}\log \L^{\,\pm1}(X(\V j))$ has a {\it short
range} representation of the type \equ(A3.5).

The SRB distribution $\m_+$ is regarded (see above) as a Gibbs state
$\lis\m_+$ with short range potential on the space $C$ of the
compatible symbolic sequences, associated with a Markov partition
$\EE$, [Si], [Ru2]. Therefore, by general large deviations properties
of short range Ising systems ([La], [El], [Ol], there is a function
$\lis\z(s)$ real analytic in $s$ for $s\in(-p^*,p^*)$ for a suitable
$p^*>0$, strictly convex and such that if $p<p^*$ and
$[p-\d,p+\d]\subset(-p^*,p^*)$ we have:

$$\fra1\t \log\lis\m_+(\{\tilde\e_\t(\V j)\in[p-\d,p+\d]\})\tende{\t\to\io}
\max_{s\in[p-\d,p+\d]}\lis\z(s)\Eqa(A3.11)$$

\0and the difference between the r.h.s.  and the l.h.s.  tends to $0$
bounded by $D\t^{\,-1}$ for a suitable constant $D$.  The function
$\lis\z(s)$ is the Legendre transform of the function $\l(\b)$ defined as:

$$\l(\b)=\lim_{\t\to\io}\fra1\t\log \ig e^{\t\b \tilde\e_\t(\V j)}\,
\lis\m_+(d\V j)\Eqa(A3.12)$$

\0\ie $\l(\b)=\max_{s\in(-p^*,p^*)}(\b s+\lis\z(s))$, where the
quantity $p^*$ can be taken $p^*=\lim_{\b\to+\io}$ $g\b^{-1}\l(\b)$
and the function $\l(\b)$ is a real analytic, [CO], strictly convex
function of $\b\in(-\io,\io)$ and $\b^{-1}\l(\b)\tende{\b\to\pm\io}\pm
p^*$, \ie it is asymptotically linear.

The above \equ(A3.11) is a "large deviations theorem" for one dimensional
spin chains with short range interactions, [La].

Hence it will be sufficient to prove the following; if
$I_{p,\d}=[p-\d,p+\d]$:

$$\fra1{\lis\h_+\t}
\log\fra{\lis\m_+(\{\tilde\e_\t(\V j)\in I_{p,\d\mp\h(\t)}\})}{\lis\m_+
(\{\tilde\e_\t(\V j)\in I_{-p,\d\pm\h(\t)}\})}\cases{<  p+\d+\h'(\t)\cr
>p-\d-\h'(\t)\cr}\Eqa(A3.13)$$

\0with $\h(\t),\h'(\t)\tende{\t\to\io}0$.
\*
\0{\it(C) Thermodynamic formalism informations.}
\*
In this section $X$ will denote a lattice interval, \ie a set of
consecutive integers $X=(x,x+1,\ldots,x+n-1)$: hence it should not be
confused with the code $X$ of \equ(A3.2).

Let $\V j_X=(j_x,j_{x+1},\ldots,j_{x+n-1})$ if
$X=(x,{x+1},\ldots,{x+n-1})$ and $n$ is odd, and call $\lis X=x+(n-1)/2$
the {\it center} of $X$. If $\V j\in C$ is an infinite spin
configuration we also denote $\V j_X$ the set of the spins with labels
$x\in X$.  The left shift of the interval $X$ will be denoted by $\th$;
\ie by the same symbol of the left shift of a (infinite) spin
configuration $\V j$.

Let $l_X(\V j_X)=l^{(n)} (j_x,j_{x+1},\ldots,j_{x+n-1})$,
and $h^+_X(\V j_X)=h^{(n)}_+(j_x,j_{x+1}, \ldots,j_{x+n-1})$ be
translation invariant,
\ie functions such that $l_{\th X}(\V j)\= l_X(\V j)$ and
$h^+_{\th X}(\V j)$$=h^+_X(\V j)$, and such that the functions $h_+(\V
j)$, see \equ(2.4), and $L(\V j)$, see \equ(A3.10), can be written
for suitably chosen constants $b_1,b_2,b,b'$:

$$\eqalign{
L(\V j)=&\sum_{\lis X= 0}l_X(\V j_X), \qquad h_+(\V j)=\sum_{\lis X=
0}h^+_X(\V
j_X)\cr
|l_X(\V j_X)|\le& b_1 e^{-b_2 n},\qquad\kern1cm |h^+_X(\V j_X)|\le b e^{-b'
n}\cr}\Eqa(A3.14)$$

\0Then $\t\tilde\e_\t (\V j)$ can be written as
$\sum_{\lis X\in[-\t/2,\t/2-1]} l_X(\V j_X)$.

Hence $\t\tilde\e_\t(\V j)$ can be approximated by $\t\tilde \e_\t^M(\V
j) ={\sum}^{(M)} l_X(\V j_X)$ where $\sum^{(M)}$ means summation over
the sets $X\subseteq[-\fra12\t-M,\fra12\t+M]$, while $\lis X$ is in
$[-\fra12\t,\fra12\t-1]$. The approximation is described by:

$$|\t\tilde\e^M_\t(\V j)-\t\tilde\e_\t(\V j)|\le b_3 e^{-b_4
M}\Eqa(A3.15)$$

\0for suitable\annota{4}{\nota One can check from \equ(A3.14),
that the constants $b_3,b_4$ can be expressed as simple functions of
$b_1,b_2$.} $b_3,b_4$ and for all $M\ge0$. Therefore if
$I_{p,\d}=[p-\d,p+\d]$ and $M=0$ we have:

$$\m_+(\{\e_\t(x)\in I_{p,\d}\})\cases{\le \lis\m_+(\{\tilde\e^0_\t\in
I_{p,\d+b_3/\t}\})\cr \ge \lis\m_+(\{\tilde\e^0_\t\in
I_{p,\d-b_3/\t}\})\cr}\Eqa(A3.16)$$

It follows from the general theory of $1$--dimensional Gibbs
distributions, [Ru2], that the $\lis\m_+$--pro\-ba\-bi\-li\-ty of a spin
configuration which coincides with $\V j_{[-\t/2,\t/2]}$ in the interval
$[-\fra12 \t,\fra12\t]$,\annota{5}{\nota \ie the spin
configurations $\V j'$ such that $j'_x=j_x$,
$x\in[-\fra12\t,\fra12\t]$.} is:

$$\fra{\Big[e^{-{\sum}^*h^+_X(\V j_X)}\Big]}
{\sum_{\V j'_{[-\t/2,\t/2]}}\Big[\cdot\Big]}\,
P(\V j_{[-\t/2,\t/2]})\Eqa(A3.17)$$

\0where $\sum^*$ denotes summation over all the $X\subseteq
[-\t/2,\t/2-1]$; the denominator is just the sum of terms like the
numerator, evaluated at a generic (compatible) spin configuration $\V
j'_{[-\t/2,\t/2]}$; finally $P$ verifies the bound, [Ru2]:

$$B_1^{-1}< P(\V j_{\,[-\t/2,\t/2]})< B_1\Eqa(A3.18)$$

\0with $B_1$ a suitable constant independent of $\V j_{\,[-\t/2,\t/2]}$
and of $\t$ ($B_1$ can be explicitly estimated in terms of $b,b'$).
Therefore from \equ(A3.16) and \equ(A3.17) we deduce for any $T\ge\t/2$:

$$\eqalign{
&\m_+(\{\e_\t(x)\in I_{p,\d}\})\le \lis\m_+(\{\tilde
\e^0_\t\in I_{p,\d+b_3/\t}\})\le\cr
&\le B_2\,\m_{T,\t}(\{\tilde\e^0_\t\in I_{p,\d+b_3/\t}\})\le
B_2\,\m_{T,\t}(\{\tilde\e_\t\in I_{p,\d+2b_3/\t}\})\cr}\Eqa(A3.19)$$

\0for some constant $B_2>0$; and likewise a lower bound is obtained by
replacing $B_2$ by $B_2^{-1}$ and $b_3$ by $-b_3$.

Then if $p<p^*$ and $I_{p,\d}\subset (-p^*,p^*)$ the set of the
rectangles $E\in\bigvee_{-T}^T S^{-k}\EE$ with center $x$ such that
$\e_\t(x)\in I_{p,\d}$ is {\it not empty}, as it follows from the strict
convexity and the asymptotic linearity of the function $\l(\b)$ in
\equ(A3.12).

We immediately deduce the lemma:
\*
\0{\it Lemma 1: the distributions $\m_+$ and $\m_{T,\t}$, $T\ge\fra12\t$,
verify:

$$\fra1{\t\lis\h_+}\log \fra{\m_+(\{\e_\t(x)\in I_{p,\d\mp 2b_3/\t}\})}
{\m_+(\{\e_\t(x)\in- I_{p,\d\pm 2b_3/\t}\})}\  \cases{
<\fra{\log B_2^2}{\t\lis\h_+}+\fra1{\t\lis\h_+}\log
\fra{\m_{T,\t}(\{\tilde\e_\t\in I_{p,\d}\})}{\m_{T,\t}(\{
\tilde\e_\t\in- I_{p,\d}\})}\cr
> -\fra{\log B_2^2}{\t\lis\h_+}+\fra1{\t\lis\h_+}\log
\fra{\m_{T,\t}(\{\tilde\e_\t\in I_{p,\d}\})}{\m_{T,\t}(\{\tilde\e_\t\in-
I_{p,\d}\})}\cr}\Eqa(A3.20)$$
\penalty10000
\0for $I_{p,\d}\subset [-p^*,p^*]$ and for $\t$ so large that
$p+\d+2b_3/\t< p^*$.}
\*
Hence \equ(A3.13) will follow if we can prove:
\*
\0{\it Lemma 2: there is a constant $\lis b$
such that the approximate SRB distribution $\m_{T,\t}$ verifies:

$$\fra1{\lis\h_+\t}\log
\fra{\m_{T,\t}(\{\tilde\e_\t\in I_{p,\d}\})}{\m_{T,\t}(\{\tilde\e_\t\in-
I_{p,\d}\})}\ \cases{\le p+\d+ \lis b/\t\cr
\ge p-\d -\lis b/\t\cr}\Eqa(A3.21)$$

\0for $\t$ large enough (so that $\d+\lis b/\t<p^*-p$) and for all
$T\ge\t/2$.}

\*
The latter lemma will be proved in \S4 and it is the only statement that
does not follow from the already existing literature.

\*
\0{\it(D) Time reversal symmetry implications}
\*

The relation \equ(A3.20) holds for any choice of the Markov partition
$\EE$. Note that if $\EE$ is a Markov pavement so is $i\EE$
(because $iS=S^{-1}i$ and $i W^u_x=W^s_{ix}$); furthermore if $\EE_1$
and $\EE_2$ are Markov pavements then $\EE=\EE_1\vee\EE_2$ is also markovian.
Therefore:

\*\0{\it Lemma 3: there exists a time reversal Markov pavement $\EE$,
\ie a Markov pavement such that $\EE=i\EE$.}

\*
This can be seen by taking any Markov pavement $\EE_0$ and setting
$\EE=\EE_0\vee i\EE_0$. Alternatively one could construct the Markov
pavement in such a way that it verifies automatically the symmetry [G2].
Since the center of a rectangle $E_\qq\in\EE_T$ can be taken to be any
point $x_\qq$ in the rectangle $E_\qq$ we can and shall suppose that the
centers of the rectangles in $\EE_T$ have been so chosen that the center
of $i E_\qq$ is $i x_\qq$, \ie the time reversal of the center $x_\qq$
of $E_\qq$.

For $\t$ large enough the set of configurations $\qq=\V j_{\,[-T,T]}$
such that $\e_\t(x)\in I_{p,\d}$ for all $x\in E_\qq$ is not
empty\annota{6}{\nota Note that $p^*=\sup_x \limsup_{\t\to+\io}
\e_\t(S^{\t/2}x)$ and let $p\in(-p^* +\d,p^*-\d)$; furthermore $\z(s)$
is smooth, hence $>-\io$, for all $|s|<p^*$.} and the ratio in
\equ(A3.21) can be written, if $x_\qq$ is the center of $E_\qq
\in\EE_T$, as:

$$\fra{\sum_{\e_\t(x_\qq)\in I_{p,\d}}
\lis\L^{\,-1}_{u,\t}(x_\qq)}{\sum_{\e_\t(x_\qq)
\in- I_{p,\d}}\lis\L^{\,-1}_{u,\t} (x_\qq)}=
\fra{\sum_{\e_\t(x_\qq)\in I_{p,\d}}
\lis\L^{\,-1}_{u,\t}(x_\qq)}{\sum_{\e_\t(x_\qq)\in
I_{p,\d}}\lis\L^{\,-1}_{u,\t}(i\,x_\qq)}\Eqa(A3.22)$$

Define $\lis\L_{s,\t}(x)$ as in \equ(A3.8) with $s$ replacing $u$:
then the time reversal symmetry implies that
$\lis\L_{u,\t}(x)=\lis\L_{s,\t}^{\,-1}(i x)$, see remark 2) following
definition (B), \S2.\annota{7}{\nota Here it is essential that
$\lis\L_{u,\t}(x)$ is the expansion of the unstable manifold between the
initial point $S^{-\t/2}x$ and the final point $S^{\t/2}x$: \ie it is a
trajectory of time length $\t$, which at its central time is in $x$.}
This permits us to change \equ(A3.22) into:

$$\fra{\sum_{\e_\t(x_\qq)\in I_{p,\d}}
\lis\L_{u,\t}^{\,-1}(x_\qq)}{\sum_{\e_\t(x_\qq)\in
I_{p,\d}}\lis\L_{s,\t}(\,x_\qq)}\ \cases{< \max_\qq
\lis\L^{\,-1}_{u,\t}(x_\qq) \lis\L^{\,-1}_{s,\t}(x_\qq) \cr>\min_\qq
\lis\L^{\,-1}_{u,\t}(x_\qq) \lis\L^{\,-1}_{s,\t}(x_\qq)\cr} \Eqa(A3.23)$$

\0where the maxima are evaluated as $\qq$ varies with $\e_\t(x_\qq)\in
I_{p,\d}$.
\\\hbox{}\kern0.3cm
By \equ(2.1) we can replace
$\lis\L^{\,-1}_{u,\t}(x)\lis\L^{\,-1}_{s,\t}(x)$ with
$\lis\L_\t^{-1}(x)B^{\pm1}$, see \equ(A3.8), \equ(2.4); thus
noting that by definition of the set of $\qq$'s in the maximum
in \equ(A3.23) we have $\fra1{\lis\h_+\t}\log \lis\L^{\,-1}_\t(x_\qq)
\in I_{p,\d}$, we see that \equ(A3.21) follows with $\lis b
=\fra1{\lis\h_+}\log B$.
\*
\0{\it Corollary: the above analysis gives us a concrete bound on the
speed at which the limits in \equ(2.6) are approached. Namely the error
has order $O(\t^{-1})$.}
\*
\0This is so because the limit \equ(A3.11) is reached at speed
$O(\t^{-1})$; furthermore the regularity of $\l(s)$ in \equ(A3.11) and the
size of $\h(\t),\h'(\t)$ and the error term in
\equ(A3.21) have all order $O(\t^{-1})$.

The above analysis proves a large deviation result for the probability
distribution $\m_+$: since $\m_+$ is a Gibbs distribution, see
\equ(A3.6), various other large deviations theorems hold for it, [DV],
[El], [Ol], but unlike the above they are not related to the time reversal
symmetry.

\*
\0{\bf Appendix A4: Heuristic proof of the local fluctuation theorem.}
\numsec=4\numfor=1\*

\0{\it(A) Markov partitions and symbolic dynamics for the chain.}
\*
The reduction of the dynamical nonequilibrium problem of a weakly
interacting chain of Anosov maps, see \S3, to a short range lattice spin
system equilibrium problem is the content of (A), (B) of this appendix,
see [Ga7]. This is an extension of the corresponding analysis in
Appendix A3 for the case of a single Anosov map: it is necessary to
discuss it again in order to exploit the short range nature of the
coupling and its weakness in order to obtain results independent on the
size $N$ of the chain.

Let $\lis \PP_0=(E^0_1,\ldots,E^0_{\NN_0})$ be a Markov partition, see
[Si], for the unperturbed ``single site'' system $(\lis M_0\times \lis
M_0, \lis S_0\times \lis S_0^{\,-1})$. Then
$\lis\PP_0^{2N+1}=\{E_\a\}$, $\a=(\r_{-N},\ldots,\r_N)$ with $E_\a=
E^0_{\r_{-N}}\times E^0_{\r_{-N+1}}\times\ldots \times E^0_{\r_N}$ is
a Markov partition of $(\lis M_0^{2(2N+1)}, S_0)$.

The perturbation, {\it if small enough}, will deform the partition
$\lis\PP_0^{2N+1}$ into a Markov partition $\PP$ for $(M,S)$ changing
only ``slightly'' the partition $\lis\PP_0^{2N+1}$. The work [PS]
shows that the above ``$\e$ small enough'' {\it mean that $\e$ has to
be chosen small but that it can be chosen $N$--independent}, as we
shall always suppose in what follows.

Under such circumstances we can establish a correspondence between
points of $M$ that have the same ``symbolic history'' (or ``symbolic
dynamics'') along $\lis\PP_0^{2N+1}$ under $S_0$ and along $\PP$ under
$S$; we shall denote it by $h$; see [PS].

The Markov partition $\lis\PP_0^{2N+1}$ for $S_0$ associates with each
point $\xx=(x_{-N},\ldots, x_N)$ a sequence $(\s_{i,j})$, $i\in
[-N,N], j\in (-\io,\io)$ of symbols so that $(\s_{i,j})_{j=-\io}^\io$
is the free symbolic dynamics of the point $x_i$. We call the first
label $i$ of $\s_{i,j}$ a ``space--label'' and the second a
``time--label''.  Not all sequences can arise as histories of points:
however (by the definition of $h$, see above) precisely the same
sequences arise as histories of points along $\PP_0$ under the free
evolution $S_0$ or along $\PP$ under the interacting evolution $S$.

The map $h$ is H\"older continuous and ``short ranged'':
$$|h(\xx)_i-h(\xx')_i|\le C\sum_j \e^{|i-j|\g'} |x_j-x'_j|^\g\Eqa(A4.1)$$
for some $\g,\g',C>0$, [PS], if $|x-y|$ denotes the distance in $\lis
M_0\times\lis M_0$ (\ie in the single site phase space).

Furthermore the code $\xx\otto\V\s$ associating with $\xx$ its
``history'' or ``symbolic dynamics'' $\V\s(\xx)$ along the partition
$\PP$ under the map $S$ is such that, fixed $j$:
$$\V\s(\xx)_i=\V\s(\xx')_i\ {\rm for}\  |i-j|\le\ell \qquad \tto\qquad
|x_j-x'_j| \le C\e^{\g \ell}\Eqa(A4.2)$$
The inverse code associating with a history $\V\s$ a point with such history
will be denoted $\xx(\V\s)$.

If $\xx=(x_{-N},\ldots, x_N)$ is coded into
$\V\s(\xx)=(\V\s_{-N},\ldots,\V\s_N)=(\s_{i,j})$, with
$i=-N,\ldots,N$, and $j\in (-\io,+\io)$, the short range property
holds also in the time direction. This means that, fixed $i_0$:
$$\s_{i,j}=\s'_{i,j}\ {\rm for}\  |i-i_0|<k, |j|<p\qquad\tto\qquad
|\xx(\V\s)_{i_0}-\xx(\V\s')_{i_0}|\le C \e^{\g k} e^
{-\k p}\Eqa(A4.3)$$
for some $\k,\g,C>0$, see lemma 1 of [PS]. The constants
$\k,\g,C,C',B,B'>0$ above and below should not be thought to be the
same even when denoted by the same symbol: however they could be {\it
a posteriori} fixed so that to equal symbols correspond equal values.

By construction the codes $\xx\otto\V\s(\xx)$ commute with time
evolution.

The sequences $(\s_{i,j})$ which arise as symbolic dynamics along
$\lis\PP_0$ under the free single site evolution of a point $x_i$ are
subject to constraints, that we call ``vertical'', imposing that
$T^0_{\s_{i,j},\s_{i,j+1}}\=1$ for all $j$, if $T^0_{\s,\s'}$ denotes
the ``compatibility matrix'' of the ``free single site evolution''
(\ie $T^0_{\s,\s'}=1$ if the $\lis S_0\times \lis S_0^{\,-1}$ image of
$E_\s$ intersects the interior of $E_{\s'}$ and $T^0_{\s,\s'}=0$
otherwise). We call the latter condition a ``compatibility condition''
for the spins in the $i$--th column.

The mixing property of the free evolution immediately implies that a
large enough power of the compatibility matrix $T^0$ has all entries
positive. This means that for each symbol $\s$ we can find
semiinfinite sequences:
$$\eqalign{
\s_B(\s)=&(\ldots,\s_{-1},\s_0\=\s), \qquad
T^0_{\s_{i-1},\s_i}=1,\quad {\rm for\ all}\ i\le0\cr
\s_T(\s)=&(\s\=\s_0,\s_1,\ldots), \qquad
T^0_{\s_{i},\s_{i+1}}=1,\quad {\rm for\ all}\ i\ge0\cr}\Eqa(A4.4)$$
and defines two functions $\s_B,\s_T$, called ``compatible
extensions'', defined on the set $\{1,\ldots,\NN_0\}$ of labels of the
single site Markov partition $\lis\PP_0$, with values in the compatible
semiinfinite sequences.

In fact there are (uncountably) many ways of performing such
compatible extensions ``from the bottom'' and ``from the top'' of the
symbol $\s$ into semiinfinite compatible sequences of symbols. We
imagine to select one pair $\s_B,\s_T$ arbitrarily, once and for all,
and call such a selection a ``choice of boundary conditions'' or ``of
extensions'', on symbolic dynamics, for reasons that should become
clear shortly.  All this seems unavoidable and it is closely parallel
to the corresponding discussion in the analysis of the simpler case of
a single Anosov system discussed in Appendix A3, see the discussion
preceding \equ(A3.8).

We shall therefore be able to ``extend in a standard way'' any finite
compatible block\annota8{\rm A block $(\s_{i,j}),\,(i,j)\in Q$, is
naturally said to be ``compatible'' if $T^0_{\s_{i,j},\s_{i,j+1}}=1$
for all $(i,j)\in Q$ such that $(i,j+1)$ is also in $Q$.} $Q$ of
spins:
$$\V\s_Q=(\s_{i,j})_{i\in L, j\in K}, \qquad L=(a-\ell,a+\ell), \
K=(b-m,b+m)\Eqa(A4.5)$$
by setting $\s_{i,j}=\s_B(\s_{i,b-n})_{b-n-j}$ for $j<b-n$ and $\s_{i,j}=
\s_T(\s_{i,b+n})_{j-b-n}$ for $j>b+n$. Here $a,b,\ell,m$ are integers.

In the free evolution there are no ``horizontal'' compatibility
constraints; hence it is always possible to extend the finite block
$\V\s_Q= (\s_{i,j})_{i\in L, j\in K}$ to a ``full spin configuration''
sequence $(\s_{i,j})_{i\in [-N,N], j\in (-\io,\io)}$, obtained by
continuing the columns in the just described standard way, using the
boundary extensions $\s_B,\s_T$, above the top and below the bottom,
into a biinfinite sequence and also by extending the spin
configuration to the right and to the left to a sequence with spatial
labels running over the full spatial range $[-N,N]$. One simply
defines $\s_{i,j}$ for $i\not\in L$ as {\it any} (but prefixed once
and for all) compatible biinfinite sequence of symbols (the same for
each column).

The allowed symbolic dynamics sequences for the free dynamics (on
$\PP_0$) and for the interacting dynamics (on $\PP$) {\it coincide}
because the free and the interacting dynamics are conjugated by the
map $h$, [PS]. Therefore the above operations make sense {\it also} if
the sequences are regarded as symbolic sequences of the interacting
dynamics, as we shall do from now on.

To conclude: given a ``block'' $\V\s_Q$ of symbols, with space--time
labels $(i,j)\in Q=L\times K$, we can associate with it a point
$\xx\in M$ whose symbolic dynamics is the above described standard
extension of $\V\s_Q$. The latter depends only on the values of
$\s_{i,j}$ for $j$ at the top or at the bottom of $Q$ and, of course, on
the boundary conditions $\s_B,\s_T$ chosen to begin with.
\*
\0{\it(B) Expansion and contraction rates.}
\*

Consider the rates of variation of the phase space volume,
$\l_0(\xx)$, or, respectively, of the surface elements of the stable
and unstable manifolds $\l_s(\xx)$ and $\l_u(\xx)$ at the point $\xx$:
they are the logarithms of the Jacobian determinants $\dpr S(\xx)$,
$\dpr_{(\a)} S(\xx)$, $\a=s,u$, where $\dpr_{(\a)} $ denotes the
Jacobian of $S$ as a map of $W^\a_\xx$ to $W^\a_{S\xx}$ where $\a=u,s$
distinguishes the unstable manifold $W^u_\xx$ of $\xx$ or the stable
manifold $W^\s_\xx$ of $\xx$:
$$\l_\a(\xx)=-\log|\det \dpr_{(\a)} S(\xx)|,\qquad \a=0,u,s\Eqa(A4.6)$$
where $\dpr_{(0)} S(\xx)\defi\dpr S(\xx)$.

A hard technical problem is to represent $\l_\a(\xx)$ in terms of the
``symbolic history'' of $\xx$ along $\PP$, \ie in terms of compatible
sequences $\V\s=(\s_{i,j})$ with $i\in (-N,N),\, j\in(-\io,\io)$.
The rates $\l_\a(\xx)$ can be expressed as:
$$\l_\a(\xx)=-\log \big| \det \fra{\dpr S}{\dpr \xx}\big|_{W^\a(\xx)}=
\sum_{L\subset [-N,N]} \tilde \d_L^{(\a)}(\xx_L)\Eqa(A4.7)$$
where $L$ is an interval in $[-N,N]$ (with $\pm(N+1)$ identified with
$\mp N$), [PS].

For $\a=0$ this can be done by noting that the matrix $J=\fra{\dpr
S}{\dpr x}$ has an almost diagonal structure:
$J(\xx)=J_0(\xx)(1+\D(\xx))$ where $J_0(\xx)$ is the Jacobian matrix of
the free motion $J_0(\xx)=\lis J_0(x_{-N})\times \lis J_0(x_{-N+1})\times
\ldots\times \lis J_0(x_N)$ if $\xx=(x_{-N},
\ldots,x_N)$ and if $D=\big(\prod_{j=-N}^N \det \lis J_0(x_j)\big)$:
$$\det J= D \cdot
e^{{\rm Tr\,}  \log(1+\D(\xx))}=
D\cdot e^{\sum_{k=1}^\io\fra{(-1)^{k-1}}k {\rm Tr\,}\D(\xx)^k}
\Eqa(A4.8)$$
which leads to \equ(A4.7) if one uses that the matrix elements
$\D_{p,q}=J_0^{-1}(\xx)\dpr_{x_p}\dpr_{x_q} J(\xx)$ are essentially
local, \ie bounded by $B\,(C\e)^{|p-q|\g}$ for some $\g,C,B>0$ (see
\equ(3.1),\equ(3.2), \equ(A4.3)).

For $\a=u,s$ \equ(A4.7) can be derived in a similar way using also
that:
\*
\0(1) the stable and unstable manifolds of $\xx$ consist of points $\yy$
which have eventually, respectively towards the future or towards the
past, the same history of $\xx$,
\\(2) they are described in a local system of coordinates around
$\xx=(\ldots,x_{-1},x_0,x_1,\ldots)$ by smooth ``short range''
functions. Suppose, in fact, that on each factor $M_0$ we introduce a
local system of coordinates $(\a,\b)$ around the point $x_i\in M_0$,
such that the unperturbed stable and unstable manifolds are described
locally by graphs $(\a,f_s(\a))$ or $(f_u(\b),\b)$.

The unperturbed stable and unstable manifolds will be smooth graphs
$(\a_i,f_s(\a_i))$ or $(f_u(\b_i),\b_i)$ with $\a_i$ varying close to
$\lis\a_i$ and $\b_i$ close to $\lis\b_i$, with $(\lis
\a_i,\lis\b_i)$ being the coordinates of $x_i$.

Fixed a point $\xx=(x_{-N},\ldots,x_N)$ with coordinates
$(\lis\a_i,\lis\b_i)_{i=-N,\ldots,N}$ the perturbed manifol\/ds of the
point $\xx$ will be described by smooth (at least $C^2$ and in fact of
any prefixed smoothness if $\e$ is sufficiently small) functions
$W^s(\V \a), W^u(\V\b)$ of $\V\a=(\a_i)_{i=-N,N}$ or of
$\V\b=(\b_i)_{i=-N,N}$ which are ``local''; \ie if $\V\a$ and $\V\a'$
agree on the sites $i-\ell,i+\ell$ or if $\V\b$ and $\V\b'$ agree on
the sites $i-\ell,i+\ell$ then:
$$\eqalign{
&||W^u(\V\b)_i-f_u(\b_i)||_{C^2}< C\e,\qquad
||W^u(\V\b)_i-W^u(\V\b')_i||_{C^2}< C\e^\ell\cr
&||W^s(\V\a)_i-f_s(\a_i)||_{C^2}< C\e,\qquad
||W^s(\V\a)_i-W^s(\V\a')_i||_{C^2}< C\e^{\ell}\cr}\Eqa(A4.9)$$
for some $C>0$, see [PS] lemmata 1,2. Here the norms in the first
column are the norms in $C^2$ as functions of the arguments $\V\b$ or
respectively $\V\a$, while the norms in the second column are $C^2$
norms evaluated (of course) after identifying the arguments of $\V\b$
(or $\aa$) and $\V\b'$ (or $\aa'$) with labels $j$ such that
$|i-j|\le\ell$.

\0(3) If we consider the dependence of the planes tangent to the
stable and unstable manifolds $W^s_\xx,\,W^u_\xx$ at $\xx$ we
find that they are H\"older continuous as functions of $\xx$:
$$|(d W^\a_\xx)_i-(d W^\a_\yy)_i|< C\,\sum_j
\e^{|i-j|\k}|x_j-y_j|^\g,\qquad \a=u,s\Eqa(A4.10)$$
where $(d W^\a_\xx)_i$ denoted the components relative to the $i$--th
coordinate of $\xx$ of the tangent plane to $W^\a_\xx$ and $C,\k,\g>0$.
\*

The above properties and the H\"older continuity \equ(A4.1),
\equ(A4.2), \equ(A4.3) imply that the ``horizontal potentials''
$\tilde\d^{(\,\a)}_L(\xx_L)$ in \equ(A4.7) are ``short ranged'':
$$|\,\tilde\d^{(\,\a)}_L(\xx_L)|\le B\,(C\e)^{(|L|-1)\g},\qquad \a=u,s
\Eqa(A4.11)$$
for some $B,C,\g >0$; we denote $|L|$ the number of points in the set
$L$.

We shall use the symbolic representation of $\xx\in M$ to express the
rates $\l^{(\a)}(\xx)$. For this purpose let $\xx=(x_i)_{i=-N,N}$ and
suppose that such $\xx$ corresponds to the symbolic dynamics sequence
$\V\s=(\V\s_j)_{j=-\io}^\io$ where
$\V\s_j=(\s_{-N,j},\ldots,\s_{N,j})$. We denote $\V\s_L$ the sequence
$\V\s_L=(\s_{i,j})_{i\in L,j=-\io,\io}$.

Then $\V\s_L$ {\it does not} determine $\xx_L$ (unless there is no
interaction, \ie $\e=0$): however the short range property, \equ(A4.3),
of the symbolic codes and of the map $h$ conjugating the free
evolution and the interacting evolution shows that, if $L'$ is a
larger interval containing $L$ and centered around $L$, then the
sequence $\V\s_{L'}$ determines each point of $\xx_L$ within an
approximation $\le(C\e)^{(|L'|-|L|)\g}$.  Hence we can define
$\widehat\d_L^{(\,\a)}(\V\s_L)$ so that:
$$\eqalign{
\tilde\d^{(\a)}_L(\xx_L)=&\sum_{L'\supset L}
\widehat\d^{(\,\a)}_{L'}(\V\s_{L'}),\qquad
|\widehat\d_L^{(\,\a)}(\V\s_L)|< B'\,(C'\e^\g)^{|L|-1}\cr
\l_\a(\xx)=&\sum_L 2^{|L|}\widehat\d_L^{(\,\a)}(\V\s_L)\cr}
\Eqa(A4.12)$$
for some $B',C',\g$. This leads to expressing $\l_\a(\xx)$ in terms of
the symbolic dynamics of $\xx$ and of the ``space--localized''
potentials $\widehat \d_L^{(\,\a)}(\V\s_L)$.

Let $Q_n=L\times K$ where $K=[-n,n]$ is a ``time--interval'' and set
$$\LL^{\,\a}_{Q_n}(\V\s_{Q_n})\defi \widehat
\d_L^{(\,\a)}([\V\s_{Q_n}])-
\widehat \d_L^{(\,\a)}([\V\s_{Q_{n-1}}])\Eqa(A4.13)$$
if $n\ge1$ and $[\V\s_{Q_n}]$ denotes a {\it standard extension} (in the
sense of \S3) of $\V\s_{Q_n}$; or just set $\LL^{\,\a}_{Q_0}\defi \widehat
\d_L^{(\,\a)}([\V\s_{Q_0}])$ for $n=0$. We define
$\LL^\a_Q(\V\s_{Q})$ for $Q=L\times K$ and $K$ not centered (\ie
$K=(a-n,a+n),\, a\ne0)$ so that it is translation invariant with
respect to space time translations (of course the horizontal
translation invariance is already implied by the above definitions and
the corresponding translation invariance of
$\widetilde\d^{(\,\a)}_L$).

The {\it remarkable property}, consequence of the H\"older continuity
of the functions in \equ(A4.6) and of the \equ(A4.3),\equ(A4.12), see
[PS], is that for some $\g,\k,B,C>0$:
$$|\LL^\a_Q(\V\s_Q)|\le B\,( C \e^\g)^i\, e^{-\k j}\Eqa(A4.14)$$
if $i,j$ are the horizontal and vertical dimensions of $Q$.

In this way we define a ``space--time local potential'' $\LL^{(\a)}_Q$
which is, by construction, translation invariant and such that, if
$\L$ denotes the box $\L=[-N,N]\times[-M,M]$ the following
representations for the rates in \equ(A4.6) hold:
$$-\log|\det\dpr_{(\,\a)}{S^{2M+1}}(S^{-M}\xx)|=\sum_{Q\subset \L}
\LL^\a_Q(\V\s_Q)+ O(|\dpr\L|)\Eqa(A4.15)$$
where $O(|\dpr\L|)$ is a ``boundary correction'' due to the fact that in
\equ(A4.15) one should really extend the sum over the $Q$'s centered at
height $\le M$ and contained in the infinite strip
$[-N,N]\times[-\io,\io]$ rather than restricting $Q$ to the region
$\L$. Hence the remainder in \equ(A4.15) can, in principle, be
explicitly written, in terms of the potentials $\LL_Q^{(\,\a)}$, in the
boundary term form usual in Statistical Mechanics of the
$2$--dimensional short range Ising model and it can be estimated to be
of $O(|\dpr\L|)$ by using \equ(A4.14).
\*

\0{\it(C) Symmetries. SRB states and f\/luctuations.}
\*

Besides the obvious translation invariance symmetry the dynamical
system has a {\it time reversal symmetry}; this is the diffeomorphism
$I$, see \equ(1.3), which {\it anticommutes} with $S$ and $S_0$:
$$I S=S^{-1} I,\qquad I S_0=S_0 I^{-1},\qquad I^2=1\Eqa(A4.16)$$
We can suppose that the Markov partition is time reversible, \ie to each
element $E_{\V\s}$ of the partition $\PP$ one can associate an element
$E_{\V \s'}=I E_{\V \s}$ which is {\it also} an element of the
partition. Here we simply use the invariance of the Markov partition
property under maps that either commute or anticommute with the
evolution $S$: hence it is not restrictive, see [Ga5],[Ga3], to suppose
that for each $\V\s$ one can define a $\V\s'$ so that $E_{\V\s'}=I
E_{\V\s}$. We shall denote such $\V\s'$ as $I\V\s$ or also $-\V\s$. For
$\e=0$, \ie for vanishing perturbation, the map $I$ will act
independently on each column of spins of $\V\s$. This property remains
valid for small perturbations; hence:
$$I\V\s=\{\s'_{i,j}\}=\{-\s_{i,-j}\}\defi -\V\s^I\Eqa(A4.17)$$
\ie time reversal simply ref\/lects the spin configuration corresponding
to a phase space point and changes ``sign'' of each spin.

The functions $\l_\a(\xx)$ and their ``potentials'' $\LL^\a_Q(\V\s_Q)$
verify, as a consequence, if $Q=[-\ell,\ell]\times[-k,k]$ is a
centered rectangle:
$$\l_{\a}(I\xx)=-\l_{\a'}(\xx),\qquad
\LL^\a_{Q}(\V\s_Q)=-\LL^{\a'}_Q(-\V\s_Q^I)\Eqa(A4.18)$$
where $\a'=s$ if $\a=u$ and $\a'=u$ if $\a=s$, $\a'=0$ if $\a=0$.  The
above symmetries will be translated into remarkable properties of
the SRB distribution.

The {\it ``local entropy production rate''} associated with the
``space like box'' $V_0=[-\ell,\ell]$ at the phase space point
$\xx=(\ldots,x_{\ell-1},x_{\ell},x_{\ell+1},\ldots)$ has been defined
in \S3 in therms of the Jacobian matrix of the map $S$. We can
likewise consider the corresponding Jacobian determinants of the
restriction of the map $S$ to the stable and unstable manifolds of
$\xx$. Such determinants will depend not only from $x_i$, $i\in V_0$,
and on the nearest neighbors variables $x_{\pm\ell}$ but {\it also} on
the other ones $x_k$ with $|k|>\ell+1$: however their dependence from
the variables with labels $|k|>\ell$ is exponentially damped as
$\e^{(|k|-\ell)\g}$, by \equ(A4.14). Thus we can define
$\h^s_{V_0},\h^u_{V_0}$ in a way completely analogous to $\h^0_{V_0}$
in \equ(3.3).

If we look at the average phase space variation rates
$\h^0_{V_0},\h^s_{V_0},\h^u_{V_0}$ between the time $-\th$ and $\th$
we can find, via a power expansion like the one in \equ(A4.8) along
the lines leading from \equ(A4.8) to \equ(A4.15), a mathematical
expression as:
$$\h^\a_{V_0}(\xx)\simeq
\sum_{Q}{}^*  \LL_Q^\a(\V\s_Q)\Eqa(A4.19)$$
where the $\sum^*_QQ$ runs over rectangles $Q$ centered at $0$--time
$Q=[a-\ell,a+\ell]\times[-k,k]$ with $[a-\ell,a+\ell]\subseteq V_0$.
This could be taken as an alternative {\it definition} of
$\h^\a_{V_0}$, as it is a rather natural expression. For our purposes,
if $V=V_0\times[-\th,\th]$, one needs to note that \equ(A4.19) holds at
least in the sense that:
$$\fra1{V_0\cdot(2\th+1)}\sum_{j=-\th}^\th \h^{(\,\a)}_{V_0}(S^j\xx)=
\fra1{V_0\cdot(2\th+1)}\sum_{Q\subset V} \LL^\a_Q(\V\s_Q)+ \fra{O(|\dpr
V|)}{|V|}\Eqa(A4.20)$$
\ie expression \equ(A4.19) can be used to compute the average local entropy
creation rate in the space--time region $V$ {\it up to boundary
corrections $O(|\dpr V|)$} (that can be neglected for the purposes of
the following discussion).

We now study the SRB distribution $\m$: denoting by $\media{F}_+$ the
average value with respect to $\m$ of the observable $F$ we can say,
see [Si], [PS], that if $\L=[-N,N]\times[-T,T]$:
$$\media{F}_+=\lim_{T\to\io} \fra{\sum_{\V\s} F(\V\s) e^{\sum_{Q\subset
\L} \LL^u_{Q}(\V\s_Q)}}{\sum_{\V\s} e^{\sum_{Q\subset
\L}\LL^u_{Q}(\V\s_Q)}}\Eqa(A4.21)$$
We want to study the properties of the f\/luctuations of:
$$p=\fra1{V \h_+}\sum_{Q\subset V}\LL^u_{Q}(\V\s_Q), \qquad
{\rm if\ \ }\h_+=\lim_{V\to\io}\fra1V\sum_{Q\subset V}\media{\LL^u_{Q}}_+
\Eqa(A4.22)$$
for which we expect a distribution of the form $\p_V(p)=
\,const\,e^{V \z(p)+O(\dpr V)}$. The SRB distribution gives
to the event that $p$ is in the interval $dp$ the probability $\p_V(p)
dp$ with:

$$\p_V(p)=\,const\, \sum_{at\ fixed\ p} e^{\sum_{Q\subset \L}
\LL^u_Q(\V\s_Q)}\Eqa(A4.23)$$
and (defining implicitly $U^u$):
$$\eqalign{
&\sum_{Q\subset\L} \LL^u_Q(\V\s_Q)=
\sum_{Q\subset V} \LL^u_Q(\V\s_Q)+
\sum_{Q\subset \L/V}\LL^u_Q(\V\s_Q)+\,O(|\dpr V|\,\k^{-1})\defi\cr
&\defi
U_V^u(\V\s_V)+U^u_{\L/V}(\V\s_{\L/V})+O(|\dpr V|\,\k^{-1})\cr}\Eqa(A4.24)$$
with $\k>0$, having used the ``short range'' properties \equ(A4.14) of
the potential.

In the sums in \equ(A4.21) we would like to sum over $\V\s_V$ and over
$\V\s_{\L/V}$ as if such spins were independent labels. This is not
possible because of the vertical compatibility constraints. However
the mixing property supposed on the free evolution implies that the
compatibility matrix $T^0$ raised to a large power $R$ has positive
entries. Hence if we leave a gap of width $R$ above and below $V$ we
can regard as independent labels the labels $\s_{i,j}$ with $i$ in the
space part $V_0$ of the region $V=V_0\times[-\th,\th]$ and
with $|j|>\th+R$, by a distance $\ge R$ above or below the region
$V$. Denoted $V+R\defi V_0\times [-\th-R,\th+R]$ remark that:
$$\sum_{Q\subset\L} \LL^u_Q(\V\s_Q)=
U_V^u(\V\s_V)+U^u_{\L/(V+R)}(\V\s_{\L/(V+R)})+
O(|\dpr V|\,(R+\k^{-1}))\Eqa(A4.25)$$
Hence, proceeding as in [GC1], we change the sum over (the dummy
label) $\V\s$ in the denominator to a sum over $-\V\s^I$ and using
$\LL^u_{Q^I}(-\V\s_Q^I)= -\LL^s_{Q}(\V\s_Q)$:
$$\fra{\p_V(p)}{\p_V(-p)}= \fra{\sum_{at\ fixed\ p} e^{\sum_{Q\subset
V}\LL^u_Q(\V\s_Q)} e^{U^u_{\L/(V+R)}(\V\s_{\L/(V+R)})}}
{\sum_{at\ fixed\ p} e^{\sum_{Q\subset
V}-\LL^s_Q(\V\s_Q)} e^{U^u_{\L/(V+R)}((-\V\s^I)_{\L/(V+R)})}}
\,e^{O(|\dpr V|)}\Eqa(A4.26)$$
with the summation being over the spin configurations in the ``whole
space--time'' $\L$, subject to the specified constraint of having the
same value for $p$, \ie the same average local entropy creation rate
in the space--time region $V$. The latter expression becomes, since
the labels $\V\s,-\V\s^I$ (respectively in the numerator and
denominator of \equ(A4.26)) are independent {\it dummy labels}:
$$\fra{\sum_{at\ fixed\ p} e^{\sum_{Q\subset
V}\LL^u_Q(\V\s_Q)} Z(\L/(V+R))}{\sum_{at\ fixed\ p} e^{\sum_{Q\subset
V}-\LL^s_Q(\V\s_Q)} Z(\L/(V+R))}\,e^{O(|\dpr V|)}\Eqa(A4.27)$$
so that by the \equ(A4.20), \equ(A4.22) and since the symmetry
relations above imply the relation $\sum_{Q\subset V}(\LL^u_Q(\V\s_Q)$
$+\LL^s_Q(\V\s_Q))= V\,\h_+\, p$, up to corrections of size $O(|\dpr
V| \k^{-1})$ we find, (note the repetition of the comparison argument
given in [GC]):
$$\fra{\p_V(p)}{\p_V(-p)}=e^{\h_+\, V\, p}\ e^{O(|\dpr V|)}\Eqa(A4.28)$$
yielding a {\it local fluctuation law}, \ie the first of
\equ(3.5). The second line of \equ(3.5) is a (simple) consequence of the above
analysis but we do not discuss it here.

\bigskip
\parindent=0pt\parskip1mm
{\bf References.}

[AA] Arnold, V., Avez, A.: {\it Ergodic problems of classical
mechanics}, Benjamin, 1966.

[BCL] Bonetto, F., Chernov, N., Lebowitz, J.: {\it }, preprint, 1998.

[BG] Bonetto, F., Gallavotti, G.: {\it Reversibility, coarse graining
and the chaoticity principle}, Communications in Mathematical Physics,
{\bf189}, 263--276, 1997.

[BGG] Bonetto, F., Gallavotti, G., Garrido, P.: {\it Chaotic
principle: an experimental test}, Physica D, {\bf 105}, 226--252, 1997

[Bo] Bowen, R.: {\it Markov partitions for Axiom A
diffeomorphisms}, American Journal of Mathematics, {\bf92}, 725--747,
1970.

[CO] Cassandro, M., Olivieri, E.: {\it Renormalization group and
analyticity in one dimension. A proof of Dobrushin's theorem},
Communications in mathematical physics, {\bf 80}, 255--269, 1981.

[DV] Donsker, M.D., Varadhan, S.R.S.: {\it Asymptotic evolution of certain
Markov processes expectations for large time}, Communications in Pure
and Applied Mathematics, {\bf 28}, 279--301, 1975; {\bf 29}, 389--461,
1976; {\bf 36}, 182--212, 1983.

[ECM2] Evans, D.J.,Cohen, E.G.D., Morriss, G.P.: {\it Probability
of second law violations in shearing steady flows}, Physical Review
Letters, {\bf 71}, 2401--2404, 1993.

[El] Ellis, R.S.: {\it Entropy, large deviations and statistical
mechanics}, New York, Springer Verlag, 1985.

[ER] Eckmann, J.P., Ruelle, D.: {\it Ergodic theory of strange
attractors}, Reviews in Modern Physics, {\bf 57}, 617--656, 1985.

[ES] Evans, Searles, : {\it Equilibrium microstates which generate
second law violating steady states}, Physical Review E, {\bf 50E},
1645--1648, 1994.

[Ga1] Gallavotti, G.: {\it Chaotic dynamics, f\/luctuations,
nonequilibrium ensembles}, Chaos, {\bf8}, 384--392, 1998.

[Ga2] Gallavotti, G.: {\it Statistical Mechanics}, Springer--Verlag,
in print.

[Ga4] Gallavotti, G.: {\it Chaotic hypothesis: Onsager reciprocity and
fluctuation--dissipa\-tion theorem}, Journal of Statistical Phys., {\bf
84}, 899--926, 1996.

[Ga3] Gallavotti, G.: {\it Reversible Anosov maps and large
deviations}, Mathematical Phy\-sics Electronic Journal, MPEJ,
(http:// mpej.unige.ch), {\bf 1}, 1--12, 1995.

[Ga5] Gallavotti, G.: {\it New methods in nonequilibrium
gases and f\/luids}, Proceedings of the conference {\sl Let's face chaos
through nonlinear dynamics}, U. of Maribor, 24 June-- 5 July 1996,
ed. M. Robnik, in print in Open Systems and Information Dynamics,
Vol. 5, 1998 (also in mp$\_$arc@ math. utexas.edu \#96-533 and
chao-dyn \#9610018).

[Ga6] Gallavotti, G.: {\it Extension of Onsager's reciprocity to large
fields and the chaotic hypothesis}, Physical Review Letters, {\bf 77},
4334--4337, 1996.

[Ga7] Gallavotti, G.: {\it A local fluctuation theorem}, in
mp$\_$arc@math. utexas. edu \#98-507.

[GC] Gallavotti, G., Cohen, E.G.D: {\it Dynamical
ensembles in nonequilibrium statistical mechanics}, Physical Review
Letters, {\bf74}, 2694--2697, 1995. And {\it Dynamical ensembles in
stationary states}, Journal of Statistical Physics, {\bf 80},
931--970, 1995.

[Gi] Gibbs, J. W.: {\it On the fundamental formula of dynamics},
American J. of Mathematics, Vol.II, 49--64, 1879. See also Volume II,
Part 2 of Gibbs' {\it  Collected Works}.

[HHP] Holian, B.L., Hoover, W.G., Posch. H.A.: {\it Resolution of
Loschmidt's paradox: the origin of irreversible behavior in reversible
atomistic dynamics}, Physical Review Letters, {\bf 59}, 10--13, 1987.

[Ku] Kurchan, J.: {\it Fluctuation theorem for stochastic dynamics},
Journal of Physics A, {\bf 31}, 3719--3729, 1998.

[LA] Levi-Civita, T., Amaldi, U.: {\it Lezioni di Meccanica
Razionale}, Zanichelli, Bo\-lo\-gna, 1927 (re\-prin\-ted 1974), volume
3.

[La]  Lanford, O.: {\it Entropy and equilibrium states in classical
statistical mechanics}, ed. A. Lenard, Lecture notes in Physics,
Springer Verlag, vol. {\bf 20}, p. 1--113, 1973.

[LLP] Lepri. S., Livi, R., Politi, A.  {\it Energy transport in
anharmonic lattices close and far from equilibrium}, preprint,
cond-mat@xyz. lanl. gov \#9709156, to appear on Physica D.

[LR] Lanford, O., Ruelle, D.: {\it Observables at infinity and states
with short range correlations in statistical mechanics},
Communications in Mathematical Physics, {\bf 13}, p. 194--215, 1969.

[LS] Lebowitz, J.L., Spohn, H.: {\it The Gallavotti-Cohen Fluctuation
Theorem for Stochastic Dynamics}, preprint, Rutgers University, June
1998.

[MP] Marchioro, C., Presutti, E.: {\it Thermodynamics of particle
systems in presence of external macroscopic fields}, Communications in
Mathematical Physics, {\bf27}, 146--154, 1972. And {\bf29}, 265--284,
1973.

[MR] Morriss, G.P., Rondoni, L.: {\it Applications of periodic orbit
theory to $N$--particle systems}, Journal of Statistical Physics, {\bf
86}, 991, 1997.

[Ol] Olla, S.: {\it Large deviations for Gibbs random fields},
Probability Theory and related fields, {\bf 77}, 343--357, 1988.

[PS] Pesin, Y.B., Sinai, Y.G.: {\it Space--time chaos in chains of
weakly inteacting hyperbolic mappimgs}, Advances in Soviet
Mathematics, {\bf3}, 165--198, 1991.

[RT] Rom-Kedar, V., Turaev, D.: {\it Big islands in dispersing
billiard--like potentials}, pre\-print, Weizmann Institute, April 2,
1998.

[Ru1] Ruelle, D.: {\it A measure associated with axiom A
attractors}, American Journal of Mathematics, {\bf98}, 619--654, 1976.
And {\it Sensitive dependence on initial conditions
and turbulent behavior of dynamical systems}, Annals of the New York
Academy of Sciences, {\bf356}, 408--416, 1978.

[Ru2] Ruelle, D.: {\it Statistical mechanics of a one dimensional
lattice gas}, Communications in Mathematical Physics, {\bf 9}, 267--278,
1968.

[Ru3] Ruelle, D.: {\it Positivity of entropy production in
nonequilibrium statistical mechanics}, Journal of Statistical Physics,
{\bf 85}, 1--25, 1996. And: {\it Entropy production in nonequilibrium
statistical mechanics}, Communications in Mathematical Physics,
{\bf189}, 365--371, 1997.

[Ru4] Ruelle, D.: {\sl Elements of differentiable dynamics
and bifurcation theory}, Academic Press, 1989.

[Si] Sinai, Y.: {\it Gibbs measures in ergodic theory}, Russian
Mathematical Surveys, {\bf 27}, 21--69, 1972 and {\it Lectures in
ergodic theory}, Lecture notes in Mathematics, Prin\-ce\-ton U. Press,
Princeton, 1977.


\Addresses
\end